\documentclass[12pt,letterpaper]{JHEP3}
\usepackage{cite}
\usepackage{epsfig}
\usepackage{amsmath}
\usepackage{amssymb}
\usepackage{url}

\title{The Multiverse Interpretation of Quantum Mechanics}

\author{Raphael Bousso$^{a,b}$ and Leonard Susskind$^{c}$\\ \\
  $^a$ Center for Theoretical Physics, Department of Physics\\
\ \  University of California, Berkeley, CA 94720, U.S.A.\\
$^b$  Lawrence Berkeley National Laboratory, Berkeley, CA 94720,
  U.S.A.\\
$^c$ Stanford Institute for Theoretical Physics and\\
\ \  Department of Physics, Stanford University, Stanford, CA 94305, U.S.A.}

\abstract{We argue that the many-worlds of quantum mechanics and the many worlds of the multiverse are the same thing, and that the multiverse is necessary to give exact operational meaning to probabilistic predictions from quantum mechanics.

  Decoherence---the modern version of wave-function collapse---is subjective in that it depends on the choice of a set of unmonitored degrees of freedom, the ``environment''. In fact decoherence is absent in the complete description of any region larger than the future light-cone of a measurement event.  However, if one restricts to the causal diamond---the largest region that can be causally probed---then the boundary of the diamond acts as a one-way membrane and thus provides a preferred choice of environment.  We argue that the global multiverse is a representation of the many-worlds (all possible decoherent causal diamond histories) in a single geometry.

  We propose that it must be possible in principle to verify quantum-mechanical predictions exactly.  This requires not only the existence of exact observables but two additional postulates: a single observer within the universe can access infinitely many identical experiments; and the outcome of each experiment must be completely definite.  In causal diamonds with finite surface area, holographic entropy bounds imply that no exact observables exist, and both postulates fail: experiments cannot be repeated infinitely many times; and decoherence is not completely irreversible, so outcomes are not definite.  We argue that our postulates can be satisfied in ``hats'' (supersymmetric multiverse regions with vanishing cosmological constant).  We propose a complementarity principle that relates the approximate observables associated with finite causal diamonds to exact observables in the hat.}

\begin{document}

\section{Introduction}
\label{sec-intro}

According to an older view of quantum mechanics, objective phenomena only occur when an observation is made and, as a result, the wave function collapses. A more modern view called decoherence considers the effects of an inaccessible environment that becomes entangled with the system of interest (including the observer). But at what point, precisely, do the virtual realities described by a quantum mechanical wave function turn into objective realities?  

This question is not about philosophy.  Without a precise form of decoherence, one cannot claim that anything really ``happened'', including the specific outcomes of experiments.  And without the ability to causally access an infinite number of precisely decohered outcomes, one cannot reliably verify the probabilistic predictions of a quantum-mechanical theory.

The purpose of this paper is to argue that these questions may be resolved by cosmology. We will offer some principles that we believe are necessary for a consistent interpretation of quantum mechanics, and we will argue that eternal inflation is the only cosmology which satisfies those principles.  There are two views of an eternally inflating multiverse: global (or parallel) vs.\ local (or series). The parallel view is like looking at a tree and seeing all its branches and twigs simultaneously. The series view is what is seen by an insect climbing from the base of the tree to a particular twig along a specific route.

In both the many-worlds interpretation of quantum mechanics and the multiverse of eternal inflation the world is viewed as an unbounded collection of parallel universes. A view that has been expressed in the past by both of us is that there is no need to add an additional layer of parallelism to the multiverse in order to interpret quantum mechanics. To put it succinctly, the many-worlds and the multiverse are the same thing~\cite{Sus05}.

\paragraph{Decoherence} Decoherence\footnote{For reviews, see~\cite{Zur03,Sch03}.  For a pedagogical introduction, see~\cite{PreNotes}.} explains why observers do not experience superpositions of macroscopically distinct quantum states, such as a superposition of an alive and a dead cat.  The key insight is that macroscopic objects tend to quickly become entangled with a large number of ``environmental'' degrees of freedom, $E$, such as thermal photons.  In practice these degrees of freedom cannot be monitored by the observer.  Whenever a subsystem $E$ is not monitored, all expectation values behave as if the remaining system is in a density matrix obtained by a partial trace over the Hilbert space of $E$.  The density matrix will be diagonal in a preferred basis determined by the nature of the interaction with the environment.  

As an example, consider an isolated quantum system $S$ with a two-dimensional Hilbert space, in the general state $a |0\rangle_S + b |1\rangle_S$.  Suppose a measurement takes place in a small spacetime region, which we may idealize as an event $M$. By this we mean that at $M$, the system $S$ interacts and becomes correlated with the pointer of an apparatus $A$:
\begin{equation}
(a |0\rangle_S + b |1\rangle_S)\otimes |0\rangle_A \to 
a\, |0 \rangle _S\otimes |0 \rangle _A + b\, |1 \rangle _S\otimes |1 \rangle _A~,
\label{eq-prem}
\end{equation}
This process is unitary and is referred to as a pre-measurement.  

We assume that the apparatus is not a closed system.   (This is certainly the case in practice for a macroscopic apparatus.)  Thus, shortly afterwards (still at $M$ in our idealization), environmental degrees of freedom $E$ scatter off of the apparatus and become entangled with it.  By unitarity, the system $SAE$ as a whole remains in a pure quantum state,\footnote{We could explicitly include an observer who becomes correlated to the apparatus through interaction with the environment, resulting in an entangled pure state of the form $a |0 \rangle _S\otimes |0 \rangle _A \otimes |0 \rangle _E \otimes |0 \rangle _O + b |1 \rangle _S\otimes |1 \rangle _A \otimes |1 \rangle _E \otimes |1 \rangle _O$.   For notational simplicity we will subsume the observer into $A$.}
\begin{equation}
|\psi \rangle = a\, |0 \rangle _S\otimes |0 \rangle _A \otimes |0 \rangle _E + b\, |1 \rangle _S\otimes |1 \rangle _A \otimes |1 \rangle _E ~.  
\label{eq-pure}
\end{equation}
We assume that the observer does not monitor the environment; therefore, he will describe the state of $SA$ by a density matrix obtained by a partial trace over the Hilbert space factor representing the environment:
\begin{equation}
\rho_{SA}={\rm Tr}_E |\psi \rangle \langle \psi| %\equiv \sum_i \mbox{}_E\langle i|
\end{equation}
This matrix is diagonal in the basis $\{ |0\rangle_S \otimes |0\rangle_A, |0\rangle_S \otimes |1\rangle_A, |1\rangle_S \otimes |0\rangle_A, |1\rangle_S \otimes |1\rangle_A\}$ of the Hilbert space of $SA$:
\begin{equation}
\rho_{SA}=\mathbf{diag}(|a|^2,0,0,|b|^2)~.
\end{equation}
This corresponds to a classical ensemble in which the pure state $|0\rangle_S \otimes |0\rangle_A$ has probability $|a|^2$ and the state $ |1\rangle_S \otimes |1\rangle_A$ has probability $|b|^2$.  

Decoherence explains the ``collapse of the wave function'' of the Copenhagen interpretation as the non-unitary evolution from a pure to a mixed state, resulting from ignorance about an entangled subsystem $E$.  It also explains the very special quantum states of macroscopic objects we experience, as the elements of the basis in which the density matrix $\rho_{SA}$ is diagonal.  This preferred basis is picked out by the apparatus configurations that scatter the environment into orthogonal states.  Because interactions are usually local in space, $\rho_{SA}$ will be diagonal with respect to a basis consisting of approximate position space eigenstates.  This explains why we perceive apparatus states $|0 \rangle_A$ (pointer up) or $|1 \rangle_A$ (pointer down), but never the equally valid basis states $|\pm\rangle_A \equiv 2^{-1/2}(|0\rangle_A \pm |1\rangle_A)$, which would correspond to superpositions of different pointer positions.

The entangled state obtained after premeasurement, Eq.~(\ref{eq-prem}) is a superposition of two unentangled pure states or ``branches''.  In each branch, the observer sees a definite outcome: $|0\rangle $ or $|1\rangle$.  This in itself does not explain, however, why a definite outcome is seen with respect to the basis $\{|0\rangle,|1\rangle\}$ rather than $\{|+\rangle, |-\rangle\}$.  Because the decomposition of Eq.~(\ref{eq-prem}) is not unique~\cite{Zur81}, the interaction with an inaccessible environment and the resulting density matrix are essential to the selection of a preferred basis of macroscopic states.

Decoherence has two important limitations: it is subjective, and it is in principle reversible.  This is a problem if we rely on decoherence for precise tests of quantum mechanical predictions.  We argue in Sec.~\ref{sec-nonhat} that causal diamonds provide a natural definition of environment in the multiverse, leading to an observer-independent notion of decoherent histories.  In Sec.~\ref{sec-hat} we argue that these histories have precise, irreversible counterparts in the ``hat''-regions of the multiverse.  We now give a more detailed overview of this paper.  

\paragraph{Outline} In Sec.~\ref{sec-nonhat} we address the first limitation of decoherence, its subjectivity.  Because coherence is never lost in the full Hilbert space $SAE$, the speed, extent, and possible outcomes of decoherence depend on the definition of the environment $E$.  This choice is made implicitly by an observer, based on practical circumstances: the environment consists of degrees of freedom that have become entangled with the system and apparatus but remain unobserved.  It is impractical, for example, to keep track of every thermal photon emitted by a table, of all of its interactions with light and air particles, and so on.  But if we did, then we would find that the entire system $SAE$ behaves as a pure state $|\psi \rangle$, which may be a ``cat state'' involving the superposition of macroscopically different matter configurations. Decoherence thus arises from the description of the world by an observer who has access only to a subsystem.  To the extent that the environment is defined by what a given observer cannot measure in practice, decoherence is subjective.

The subjectivity of decoherence is not a problem as long as we are content to explain our own experience, i.e., that of an observer immersed in a much larger system.  But the lack of any environment implies that decoherence cannot occur in a complete unitary description of the whole universe.  It is possible that no such description exists for our universe.  In Sec.~\ref{sec-future} we will argue, however, that causality places restrictions on decoherence in much smaller regions, in which the applicability of unitary quantum-mechanical evolution seems beyond doubt.  

In Sec.~\ref{sec-global}, we apply our analysis of decoherence and causality to eternal inflation.  We will obtain a straightforward but perhaps surprising consequence: in a global description of an eternally inflating spacetime, decoherence cannot occur; so it is inconsistent to imagine that pocket universes or vacuum bubbles nucleate at particular locations and times.   In Sec.~\ref{sec-simplicio}, we discuss a number of attempts to rescue a unitary global description and conclude that they do not succeed.

In Sec.~\ref{sec-patch}, we review the ``causal diamond'' description of the multiverse.  The causal diamond is the largest spacetime region that can be causally probed, and it can be thought of as the past light-cone from a point on the future conformal boundary.  We argue that the causal diamond description leads to a natural, observer-independent choice of environment: because its boundary is light-like, it acts as a one-way membrane and degrees of freedom that leave the diamond do not return except in very special cases.  These degrees of freedom can be traced over, leading to a branching tree of causal diamond histories.

Next, we turn to the question of whether the global picture of the multiverse can be recovered from the decoherent causal diamonds.  In Sec.~\ref{sec-dual}, we review a known duality between the causal diamond and a particular foliation of the global geometry known as light-cone time: both give the same probabilities.  This duality took the standard global picture as a starting point, but in Sec.~\ref{sec-everett}, we reinterpret it as a way of reconstructing the global viewpoint from the local one.  If the causal diamond histories are the many-worlds, this construction shows that the multiverse is the many-worlds pieced together in a single geometry.

In Sec.~\ref{sec-hat} we turn to the second limitation associated with decoherence, its reversibility.  Consider a causal diamond with finite maximal boundary area $A_{\rm max}$.  Entropy bounds imply that such diamonds can be described by a Hilbert space with finite dimension no greater than $\exp(A_{\rm max}/2)$~\cite{CEB2,Bou00a}.\footnote{This point has long been emphasized by Banks and Fischler~\cite{Ban00,Fis00b,Ban01}.}  This means that no observables in such diamonds can be defined with infinite precision.  In Sec.~\ref{sec-reverse} and \ref{sec-limitation}, we will discuss another implication of this finiteness: there is a tiny but nonzero probability that decoherence will be undone.  This means that the decoherent histories of causal diamonds, and the reconstruction of a global spacetime from such diamonds, is not completely exact.  

No matter how good an approximation is, it is important to understand the precise statement that it is an approximation to.  In Sec.~\ref{sec-sagredo}, we will develop two postulates that should be satisfied by a fundamental quantum-mechanical theory if decoherence is to be sharp and the associated probabilities operationally meaningful: decoherence must be irreversible, and it must occur infinitely many times for a given experiment in a single causally connected region.

The string landscape contains supersymmetric vacua with exactly vanishing cosmological constant.  Causal diamonds which enter such vacua have infinite boundary area at late times.  We argue in Sec.~\ref{sec-hats} that in these ``hat'' regions, all our postulates can be satisfied.  Exact observables can exist and decoherence by the mechanism of Sec.~\ref{sec-patch} can be truly irreversible.  Moreover, because the hat is a spatially open, statistically homogeneous universe, anything that happens in the hat will happen infinitely many times.

In Sec.~\ref{sec-complementarity} we review black hole complementarity, and we conjecture an analogous ``hat complementarity'' for the multiverse.  It ensures that the approximate observables and approximate decoherence of causal diamonds with finite area (Sec.~\ref{sec-patch}) have precise counterparts in the hat.  In Sec.~\ref{sec-ct} we propose a relation between the global multiverse reconstruction of Sec.~\ref{sec-everett}, and the Census Taker cutoff~\cite{Sus07} on the hat geometry.

Two interesting papers have recently explored relations between the many-worlds interpretation and the multiverse~\cite{AguTeg10,Nom11}.  The present work differs substantially in a number of aspects.  Among them is the notion that causal diamonds provide a preferred environment for decoherence, our view of the global multiverse as a patchwork of decoherent causal diamonds, our postulates requiring irreversible entanglement and infinite repetition, and the associated role we ascribe to hat regions of the multiverse.

\section{Building the multiverse from the many worlds of causal diamonds}
\label{sec-nonhat}

\subsection{Decoherence and causality}
\label{sec-future}

\begin{figure*}
\begin{center}
\includegraphics[scale = .4]{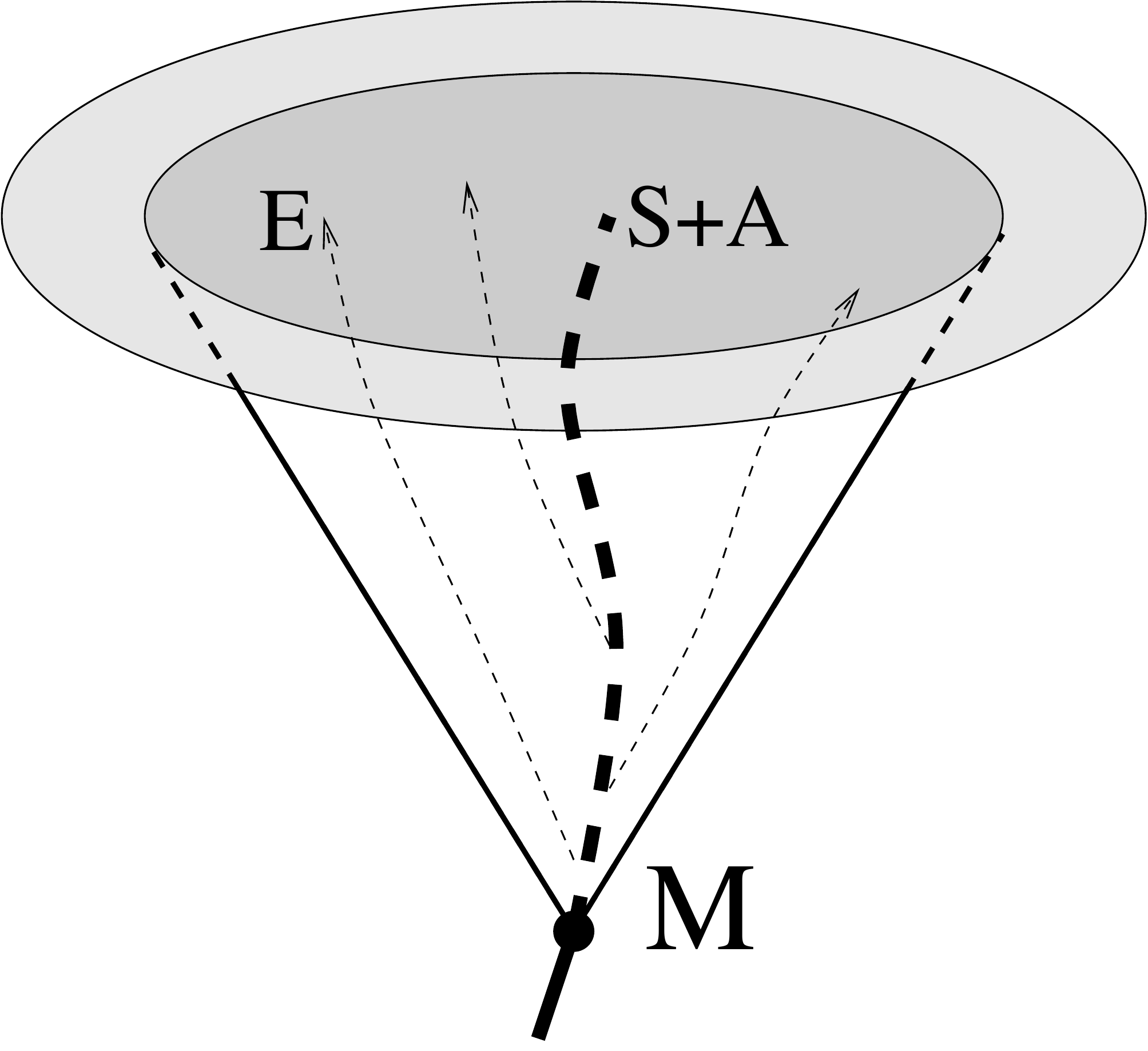}
\end{center}
\caption{Decoherence and causality. At the event $M$, a macroscopic apparatus $A$ becomes correlated with a quantum system $S$.  Thereafter, environmental degrees of freedom $E$ interact with the apparatus.  In practice, an observer viewing the apparatus is ignorant of the exact state of the environment and so must trace over this Hilbert space factor.  This results in a mixed state which is diagonal in a particular ``pointer'' basis picked out by the interaction between $E$ and $A$.  The state of the full system $SAE$, however, remains pure.  In particular, decoherence does not take place, and no preferred bases arises, in a complete description of any region larger than the future lightcone of $M$.}
\label{fig-futconedec}
\end{figure*} 

The decoherence mechanism reviewed above relies on ignoring the degrees of freedom that a given observer fails to monitor, which is fine if our goal is to explain the experiences of that observer.  But this subjective viewpoint clashes with the impersonal, unitary description of large spacetime regions---the viewpoint usually adopted in cosmology.   We are free, of course, to pick any subsystem and trace over it.  But the outcome will depend on this choice.  The usual choices implicitly involve locality but not in a unique way. 

For example, we might choose $S$ to be an electron and $E$ to be the inanimate laboratory. The system's wave function collapses when the electron becomes entangled with some detector. But we may also include in $S$ everything out to the edge of the solar system.  The environment is whatever is out beyond the orbit of Pluto. In that case the collapse of the system wavefunction cannot take place until a photon from the detector has passed Pluto's orbit. This would take about a five hours during which the system wavefunction is coherent.

In particular, {\em decoherence cannot occur in the complete quantum description of any region larger than the future light-cone of the measurement event $M$} (Fig.~\ref{fig-futconedec}).  All environmental degrees of freedom that could have become entangled with the apparatus since the measurement took place must lie within this lightcone and hence are included, not traced over, in a complete description of the state.  An example of such a region is the whole universe, i.e., any Cauchy surface to the future of $M$.  But at least at sufficiently early times, the future light-cone of $M$ will be much smaller than the whole universe.  Already on this scale, the system $SAE$ will be coherent.

In our earlier example, suppose that we measure the spin of an electron that is initially prepared in a superposition of spin-up and spin-down, $a |0\rangle_S + b |1\rangle_S$, resulting in the state $|\psi \rangle$ of Eq.~(\ref{eq-pure}).  A complete description of the solar system (defined as the interior of a sphere the size of Pluto's orbit, with a light-crossing time of about 10 hours) by a local quantum field theory contains every particle that could possibly have interacted with the apparatus after the measurement, for about 5 hours.  This description would maintain the coherence of the macroscopic superpositions implicit in the state $|\psi \rangle$, such as apparatus-up with apparatus-down, until the first photons that are entangled with the apparatus leave the solar system.

Of course, a detailed knowledge of the quantum state in such large regions is unavailable to a realistic observer.  (Indeed, if the region is larger than a cosmological event horizon, then its quantum state is cannot be probed at all without violating causality.)  Yet, our theoretical description of matter fields in spacetime retains, in principle, all degrees of freedom and full coherence of the quantum state.   In theoretical cosmology, this can lead to inconsistencies, if we describe regions that are larger than the future light-cones of events that we nevertheless treat as decohered.  We now consider an important example.

\subsection{Failure to decohere: A problem with the global multiverse} 
\label{sec-global}

The above analysis undermines what we will call the ``standard global picture'' of an eternally inflating spacetime. Consider an effective potential containing at least one inflating false vacuum, i.e. a metastable de~Sitter vacuum with decay rate much less than one decay per Hubble volume and Hubble time.  We will also assume that there is at least one terminal vacuum, with nonpositive cosmological constant.  (The string theory landscape is believed to have at least $10^{100{\rm 's}}$ of vacua of both types~\cite{BP,KKLT,Sus03,DenDou04b}.)   

According to the standard description of eternal inflation, an inflating vacuum nucleates bubble-universes in a statistical manner similar to the way superheated water nucleates bubbles of steam. That process is described by classical stochastic production of bubbles which occurs randomly but the randomness is classical. The bubbles nucleate at definite locations and  coherent quantum mechanical interference plays no role. The conventional description of eternal inflation similarly based on classical stochastic processes. However, this  picture is not consistent with  a complete quantum-mechanical description of a global region of the multiverse.

\begin{figure*}
\begin{center}
\includegraphics[scale = .35]{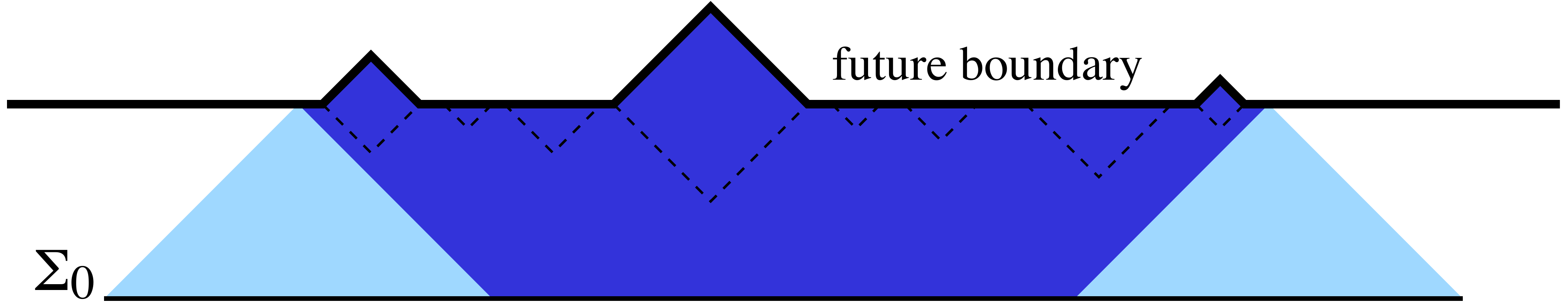}
\end{center}
\caption{The future domain of dependence, $D(\Sigma_0)$, (light or dark shaded) is the spacetime region that can be predicted from data on the timeslice $\Sigma_0$.   If the future conformal boundary contains spacelike portions, as in eternal inflation or inside a black hole, then the future light-cones of events in the dark shaded region remain entirely within $D(\Sigma_0)$.  Pure quantum states do not decohere in this region, in a complete description of $D(\Sigma_0)$.  This is true even for states that involve macroscopic superpositions, such as the locations of pocket universes in eternal inflation (dashed lines), calling into question the self-consistency of the global picture of eternal inflation.}
\label{fig-dsigma}
\end{figure*}

To explain why this is so, consider the future domain of dependence $D(\Sigma_0)$ of a sufficiently large hypersurface $\Sigma_0$, which need not be a Cauchy surface.  $D(\Sigma_0)$ consists of all events that can be predicted from data on $\Sigma_0$; see Fig.~\ref{fig-dsigma}.  If $\Sigma_0$ contains sufficiently large and long-lived metastable de~Sitter regions, then bubbles of vacua of lower energy do not consume the parent de~Sitter vacua in which they nucleate~\cite{GutWei83}.   Hence, the de~Sitter vacua are said to inflate eternally, producing an unbounded number of bubble universes.  The resulting spacetime is said to have the structure shown in the conformal diagram in Fig.~\ref{fig-dsigma}, with bubbles nucleating at definite spacetime events.  The future conformal boundary is spacelike in regions with negative cosmological constant, corresponding to a local big crunch.  The boundary contains null ``hats'' in regions occupied by vacua with $\Lambda=0$.  

But this picture does not arise in a complete quantum description of $D(\Sigma_0)$.  The future light-cones of events at late times are much smaller than $D(\Sigma_0)$.  In any state that describes the entire spacetime region $D(\Sigma_0)$, decoherence can only take place at the margin of $D(\Sigma_0)$ (shown light shaded in Fig.~\ref{fig-dsigma}), in the region from which particles can escape into the complement of $D(\Sigma_0)$ in the full spacetime.  No decoherence can take place in the infinite spacetime region defined by the past domain of dependence of the future boundary of $D(\Sigma_0)$.  In this region, quantum evolution remains coherent even if it results in the superposition of macroscopically distinct matter or spacetime configurations. 

An important example is the superposition of vacuum decays taking place at different places.  Without decoherence, it makes no sense to say that bubbles nucleate at particular times and locations; rather, a wavefunction with initial support only in the parent vacuum develops into a superposition of parent and daughter vacua.  Bubbles nucleating at all places at times are ``quantum superimposed''.  With the gravitational backreaction included, the metric, too, would remain in a quantum-mechanical superposition.  This contradicts the standard global picture of eternal inflation, in which domain walls, vacua, and the spacetime metric take on definite values, as if drawn from a density matrix obtained by tracing over some degrees of freedom, and as if the interaction with these degrees of freedom had picked out a preferred basis that eliminates the quantum superposition of bubbles and vacua.

Let us quickly get rid of one red herring:  Can the standard geometry of eternal inflation be recovered by using so-called semi-classical gravity in which the metric is sourced  by the expectation value of the energy-momentum tensor,
\begin{equation}
G_{\mu\nu}=8\pi \langle T_{\mu\nu} \rangle~?
\label{eq-scg}
\end{equation}
This does not work because the matter quantum fields would still remain coherent.  At the level of the quantum fields, the wavefunction initially has support only in the false vacuum.  Over time it evolves to a superposition of the false vacuum (with decreasing amplitude), with the true vacuum (with increasing amplitude), plus a superposition of expanding and colliding domain walls.  This state is quite complicated but the expectation value of its stress tensor should remain spatially homogeneous if it was so initially.  The net effect, over time, would be a continuous conversion of vacuum energy into ordinary matter or radiation (from the collision of bubbles and motion of the scalar field).  By Eq.~(\ref{eq-scg}), the geometry spacetime would respond to the homogeneous glide of the vacuum energy to negative values.  This would result in a global crunch after finite time, in stark contrast to the standard picture of global eternal inflation.  In any case, it seems implausible that semi-classical gravity should apply in a situation in which coherent branches of the wavefunction have radically different gravitational back-reaction.  The AdS/CFT correspondence provides an explicit counterexample, since the superposition of two CFT states that correspond to different classical geometries must correspond to a quantum superposition of the two metrics.

The conclusion that we come to from these considerations is not that the global multiverse is meaningless, but that the parallel view should not be implemented by unitary quantum mechanics.  But is there an alternative?  Can the standard global picture be recovered by considering an observer who has access only to some of the degrees of freedom of the multiverse, and appealing to decoherence?  We debate this question in the following section.

\subsection{Simplicio's proposal}
\label{sec-simplicio}

{\em Simplicio and Sagredo have studied Sections~\ref{sec-future} and \ref{sec-global}, supplied to them by Salviati.  They meet at Sagredo's house for a discussion.}

{\sc Simplicio:} You have convinced me that a complete description of eternal inflation by unitary quantum evolution on global slices will not lead to a picture in which bubbles form at definite places and times.  But all I need is an observer somewhere!  Then I can take this observer's point of view and trace over the degrees of freedom that are inaccessible to him.  This decoheres events, such as bubble nucleations, in the entire global multiverse.  It actually helps that some regions are causally disconnected from the observer: this makes his environment---the degrees of freedom he fails to access---really huge.

{\sc Sagredo:} An interesting idea.  But you seem to include everything outside the observer's horizon region in what you call the environment.  Once you trace over it, it is gone from your description and you could not possibly recover a global spacetime.

{\sc Simplicio:} Your objection is valid, but it also shows me how to fix my proposal.  The observer should only trace over environmental degrees in his own horizon.  Decoherence is very efficient, so this should suffice.

{\sc Sagredo:} I wonder what would happen if there were two observers in widely separated regions.  If one observer's environment is enough to decohere the whole universe, which one should we pick?

{\sc Simplicio:} I have not done a calculation but it seems to me that it shouldn't matter. The outcome of an experiment by one of the observers should be the same, no matter which observer's environment I trace over.   That is certainly how it works when you and I both stare at the same apparatus.

{\sc Sagredo:} Something is different about the multiverse.  When you and I both observe Salviati, we all become correlated by interactions with a common environment.  But how does an observer in one horizon volume become correlated with an object in another horizon volume far away?

{\sc Salviati:} Sagredo, you hit the nail on the head.  Decoherence requires the interaction of environmental degrees of freedom with the apparatus and the observer.  This entangles them, and it leads to a density matrix once the environment is ignored by the observer.  But an observer cannot have interacted with degrees of freedom that were never present in his past light-cone.

\begin{figure*}
\begin{center}
\includegraphics[scale = .4]{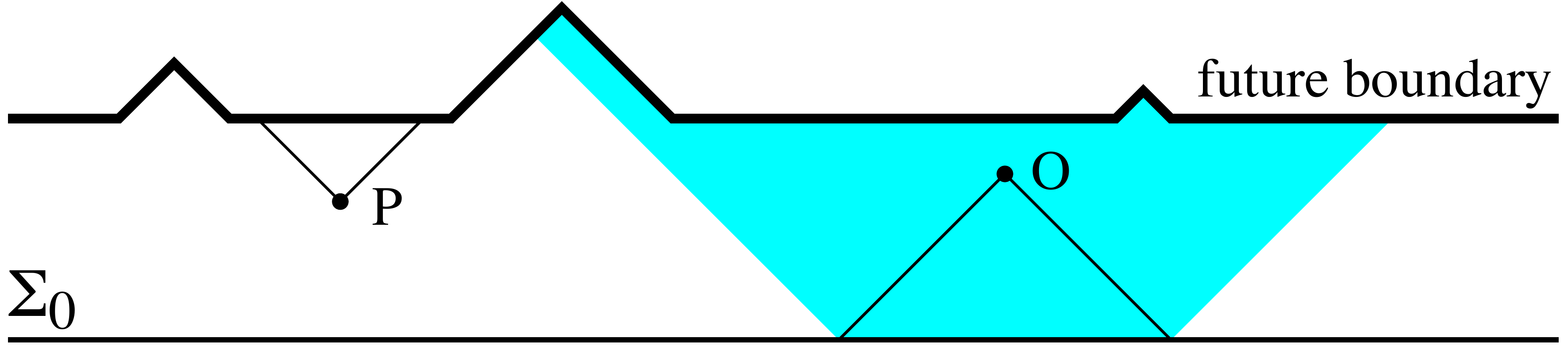}
\end{center}
\caption{Environmental degrees of freedom entangled with an observer at $O$ remain within the causal future of the causal past of $O$, $J^+[J^-(O)]$ (cyan/shaded).  They are not entangled with distant regions of the multiverse.  Tracing over them will not lead to decoherence of a bubble nucleated at $P$, for example, and hence will fail to reproduce the standard global picture of eternal inflation.}
\label{fig-sagredo}
\end{figure*} 
{\sc Sagredo:} Thank you for articulating so clearly what to me was only a vague concern.  Simplicio, you look puzzled, so let me summarize our objection in my own words.  You proposed a method for obtaining the standard global picture of eternal inflation: you claim that we need only identify an arbitrary observer in the multiverse and trace over his environment.  If we defined the environment as {\em all\/} degrees of freedom the observer fails to monitor, then it would include the causally disconnected regions outside his horizon.  With this definition, these regions will disappear entirely from your description, in conflict with the global picture.   So we agreed to define the environment as the degrees of freedom that have interacted with the observer and which he cannot access in practice.  But in this case, the environment includes {\em no\/} degrees of freedom outside the causal future of the observer's causal past.  I have drawn this region in Fig.~\ref{fig-sagredo}.   But tracing over an environment can only decohere degrees of freedom that it is entangled with.  In this case, it can decohere some events that lie in the observer's past light-cone.  But it cannot affect quantum coherence in far-away horizon regions, because the environment you have picked is not entangled with these regions.  In those regions, bubble walls and vacua will remain in superposition, which again conflicts with the standard global picture of eternal inflation.

{\sc Simplicio:} I see that my idea still has some problems.  I will need to identify more than one observer-environment pair.  In fact,  if I wish to preserve the global picture of the multiverse, I will have to assume that an observer is present in every horizon volume, at all times!   Otherwise, there will be horizon regions where no one is around to decide which degrees of freedom are hard to keep track of, so there is no way to identify and trace over an environment.  In such regions, bubbles would not form at particular places and times, in conflict with the standard global picture. 

{\sc Sagredo:} But this assumption is clearly violated in many landscape models.  Most de Sitter vacua have large cosmological constant, so that a single horizon volume is too small to contain the large number of degrees of freedom required for an observer. And regions with small vacuum energy may be very long lived, so the corresponding bubbles contain many horizon volumes that are completely empty.  I'm afraid, Simplicio, that your efforts to rescue the global multiverse are destined to fail.

{\sc Salviati:} Why don't we back up a little and return to Simplicio's initial suggestion.  Sagredo, you objected that everything outside an observer's horizon would naturally be part of his environment and would be gone from our description if we trace over it...

{\sc Sagredo:} ...which means that the whole global description would be gone...

{\sc Salviati:} ...but why is that a problem?  No observer inside the universe can ever see more than what is in their past light-cone at late times, or more precisely, in their causal diamond.  We may not be able to recover the global picture by tracing over the region behind an observer's horizon, but the same procedure might well achieve decoherence in the region the observer can actually access.  In fact, we don't even need an actual observer: we can get decoherence by tracing over degrees of freedom that leave the causal horizon of any worldline!  This will allow us to say that a bubble formed in one place and not another. So why don't we give up on the global description for a moment.  Later on, we can check whether a global picture can be recovered in some way from the decoherent causal diamonds.

{\em Salviati hands out Sections~\ref{sec-patch}--\ref{sec-everett}.}

\subsection{Objective decoherence from the causal diamond}
\label{sec-patch}

If Hawking radiation contains the full information about the quantum state of a star that collapsed to form a black hole, then there is an apparent paradox.  The star is located inside the black hole at spacelike separation from the Hawking cloud; hence, two copies of the original quantum information are present simultaneously.  The xeroxing of quantum information, however, conflicts with the linearity of quantum mechanics~\cite{WooZur82}.  The paradox is resolved by ``black hole complementarity''~\cite{SusTho93}.   By causality, no observer can see both copies of the information.  A theory of everything should be able to describe any experiment that can actually be performed by some observer in the universe, but it need not describe the global viewpoint of a ``superobserver'', who sees both the interior and the exterior of a black hole.  Evidently, the global description is inconsistent and must be rejected.
\begin{figure*}
\begin{center}
\includegraphics[scale = .5]{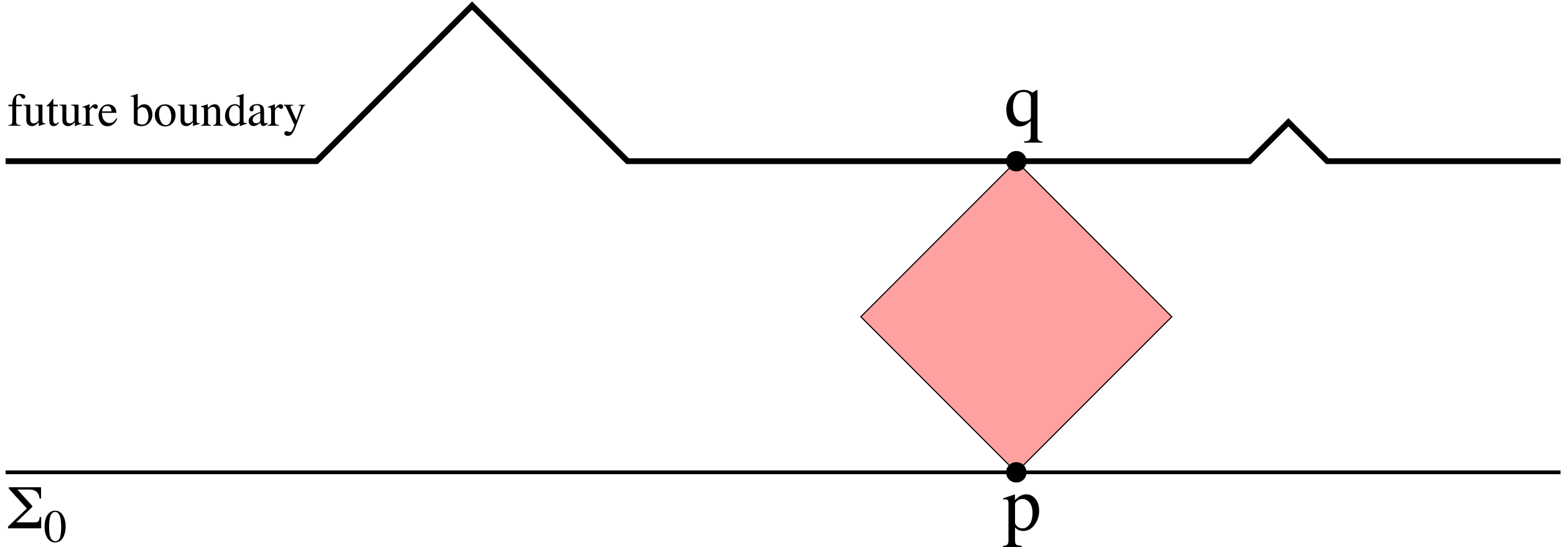}
\end{center}
\caption{The causal diamond (pink/shaded) spanned by two events $p$ and $q$ is the set of points that lie on causal curves from $p$ to $q$.  $p$ is called the origin and $q$ the tip of the causal diamond.  In the example shown, $p$ lies on the initial surface and $q$ on the future conformal boundary of the spacetime.  The causal diamond is largest spacetime region that can be causally probed by an observer travelling from $p$ to $q$.}
\label{fig-cdpq}
\end{figure*} 

If the global viewpoint fails in a black hole geometry, then it must be abandoned in any spacetime.  Hence, it is useful to characterize generally what spacetime regions can be causally probed.  An experiment beginning at a spacetime event $p$ and ending at the event $q$ in the future of $p$ can probe the {\em causal diamond\/} $I^+(p)\cap I^-(q)$ (Fig.~\ref{fig-cdpq}).  By starting earlier or finishing later, the causal diamond can be enlarged.  In spacetimes with a spacelike future boundary, such as black holes and many cosmological solutions, the global universe is much larger than any causal diamond it contains.  Here we will be interested in diamonds that are as large as possible, in the sense that $p$ and $q$ correspond to the past and future endpoints of an inextendible worldline.

We will now argue that the causal diamond can play a useful role in making decoherence more objective.  Our discussion will be completely general, though for concreteness it can be useful to think of causal diamonds in a landscape, which start in a de~Sitter vacuum and end up, after a number of decays, in a crunching $\Lambda<0$ vacuum.

\begin{figure*}
\begin{center}
\includegraphics[scale = .3]{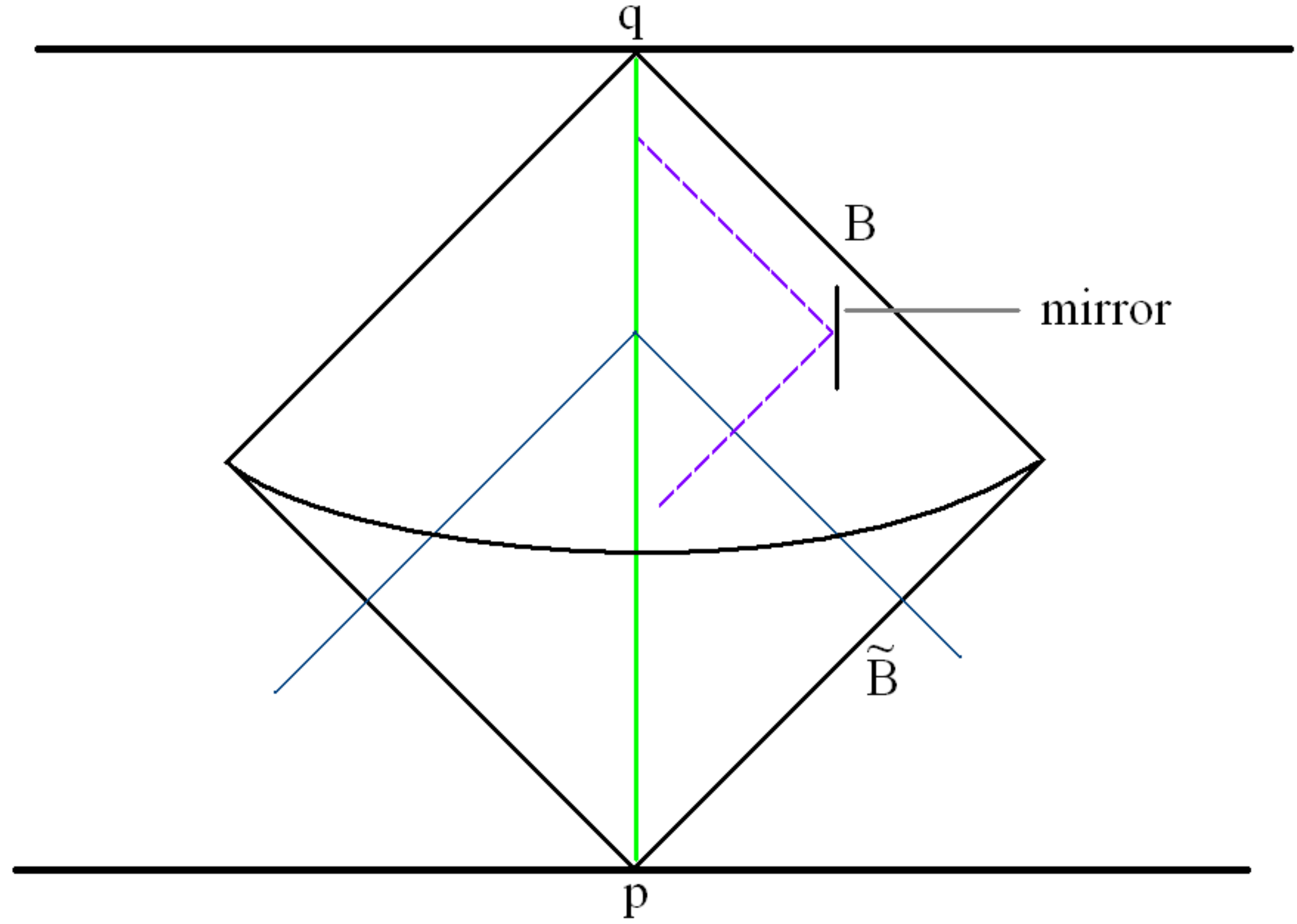}
\end{center}
\caption{Causal diamond spanned by the world-line (green) of an observer.  Environmental degrees of freedom (purple dashed line) that leave the observer's past light-cone (blue) at some finite time can be recovered using mirrors.}
\label{fig-mirror}
\end{figure*}

Consider a causal diamond, $C$, with future boundary $B$ and past boundary $\tilde B$, as shown in Fig.~\ref{fig-mirror}.  For simplicity, suppose that the initial state on $\tilde B$ is pure.  Matter degrees of freedom that leave the diamond by crossing $B$ become inaccessible to any experiment within $C$, by causality.  Therefore they {\em must\/} be traced over.

In practice, there will be many other degrees of freedom that an observer fails to control, including most degrees of freedom that have exited his past light-cone at any finite time along his worldline.  But such degrees of freedom can be reflected by mirrors, or in some other way change their direction of motion back towards the observer (Fig.~\ref{fig-mirror}).  Thus, at least in principle,  the observer could later be brought into contact again with any degrees of freedom that remain within the causal diamond $C$, restoring coherence.  Also, the observer at finite time has not had an opportunity to observe degrees of freedom coming from the portion outside his past lightcone on $\tilde B$; but those he might observe by waiting longer.  Hence, we will be interested only in degrees of freedom that leave $C$ by crossing the boundary $B$.  

The boundary $B$ may contain components that are the event horizons of black holes.  If black hole evaporation is unitary, then such degrees of freedom will be returned to the interior of the causal diamond in the form of Hawking radiation.  We can treat this formally by replacing the black hole with a membrane that contains the relevant degrees of freedom at the stretched horizon and releases them as it shrinks to zero size~\cite{SusTho93}.  However, we insist that degrees of freedom crossing the outermost component of $B$ (which corresponds to the event horizon in de Sitter universes) are traced over.  It does not matter for this purpose whether we regard these degrees of freedom as being absorbed by the boundary or as crossing through the boundary, as long as we assume that they are inaccessible to any experiment performed within $C$.  This assumption seems reasonable, since there is no compelling argument that the unitarity evaporation of black holes should extend to cosmological event horizons. Indeed, it is unclear how the statement of unitarity would be formulated in that context.  (A contrary viewpoint, which ascribes unitarity even to non-Killing horizons, is explored in Ref.~\cite{Sus07}.)
\begin{figure*}
\begin{center}
\includegraphics[scale = .4]{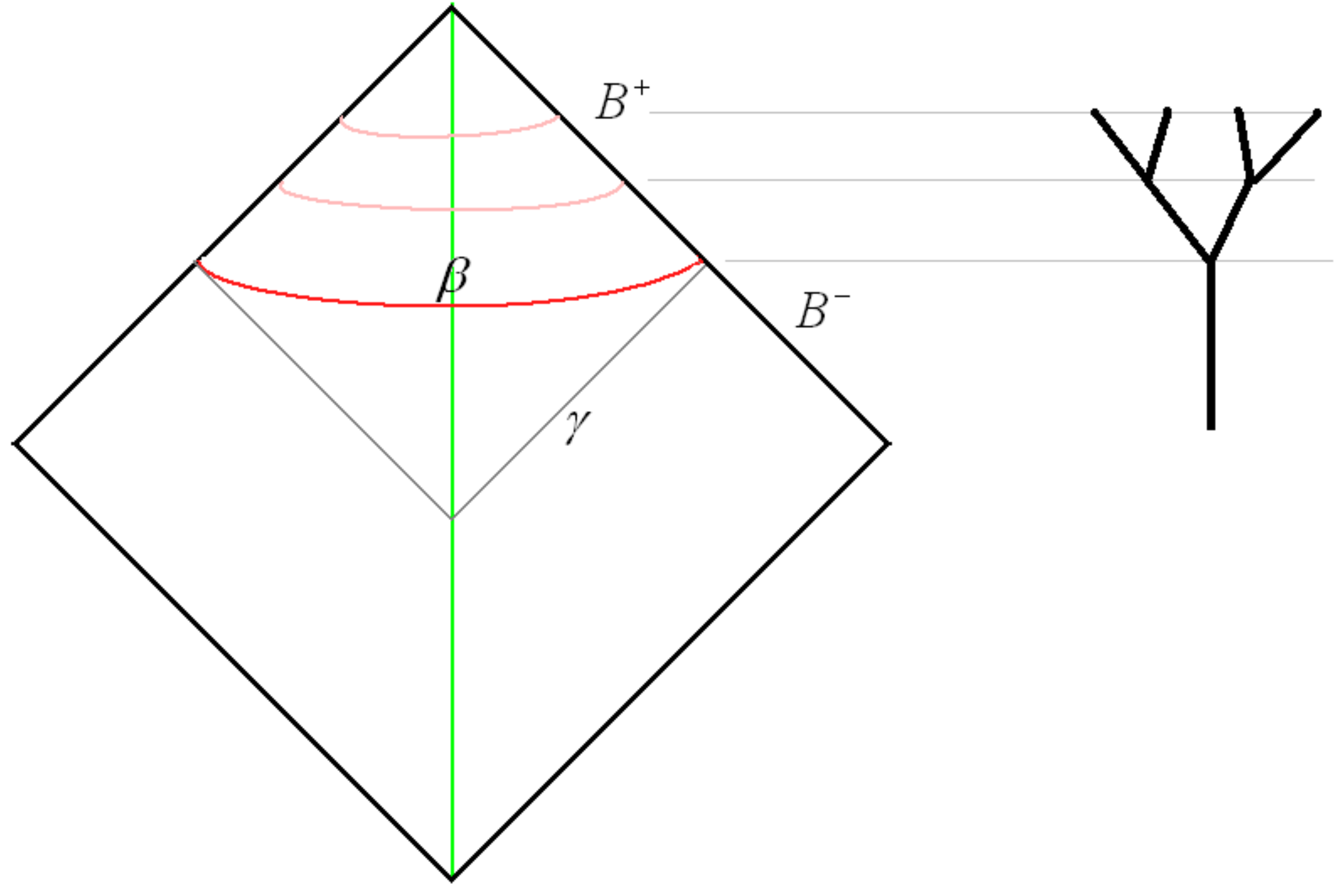}
\end{center}
\caption{The surface $\beta$ divides the future boundary of the causal diamond into two portions $B_\pm$.  Degrees of freedom that passed through $B_-$ are forever inaccessible from within the diamond.  Tracing over them defines a density matrix at the time $\gamma$.  The pure states that diagonalize this matrix can be represented as branches.  As more degrees of freedom leave the causal diamond, a branching tree is generated that represents all possible decoherent histories within the diamond.}
\label{fig-beta}
\end{figure*} 

The boundary $B$ is a null hypersurface.  Consider a cross-section $\beta$ of $B$, i.e., a spacelike two-dimensional surface that divides $B$ into two portions: the upper portion, $B_+$, which contains the tip of the causal diamond, and the lower portion $B_-$.  We may trace over degrees of freedom on $B_-$; this corresponds to the matter that has left the causal diamond by the time $\beta$ and hence has become inaccessible from within the diamond.  Thus we obtain a density matrix $\rho(\beta)$ on the portion $B_+$.  Assuming unitary evolution of closed systems, the same density matrix also determines the state on any spacelike surface bounded by $\beta$; and it determines the state on the portion of the boundary of the past of $\beta$ that lies within $C$, $\gamma$.  Note that $\gamma$ is a null hypersurface.  In fact, $\gamma$ can be chosen to be a future lightcone from an event inside $C$ (more precisely, the portion of that light-cone that lies within $C$); the intersection of $\gamma$ with $B$ then defines $\beta$.

A useful way of thinking about $\rho(\beta)$ is as follows. The boundary of the causal past of $\beta$ consists of two portions, $\gamma$ and $B_-$.  The degrees of freedom that cross $B_-$ are analogous to the environment in the usual discussion of decoherence, except in that they are inaccessible from within the causal diamond $C$ not just in practice but in principle.  The remaining degrees of freedom in the past of $\beta$ cross through $\gamma$ and thus stay inside the causal diamond. They are analogous to the system and apparatus, which are now in one of the states represented in the density matrix $\rho(\beta)$.  A measurement is an interaction between degrees of freedom that later pass through $\gamma$ and degrees of freedom that later pass through $B_-$. The basis in which $\rho(\beta)$ is diagonal consists of the different pure states that could result from the outcome of measurements in the causal past of $\beta$.  

We can now go further and consider foliations of the boundary $B$.  Each member of the foliation is a two-dimensional spatial surface $\beta$ dividing the boundary into portions $B_+$ and $B_-$.  We can regard $\beta$ as a time variable.  For example, any a foliation of the causal diamond $C$ into three-dimensional spacelike hypersurfaces of equal time $\beta$ will induce a foliation of the boundary $B$ into two-dimensional spacelike surfaces.  Another example, on which we will focus, is shown in Fig.~\ref{fig-beta}: consider an arbitrary time-like worldline that ends at the tip of the causal diamond.  Now construct the future light-cone from every point on the worldline.  This will induce a foliation of $B$ into slices $\beta$.  It is convenient to identify $\beta$ with the proper time along the worldline.

The sequence of density matrices $\rho(\beta_1), \rho(\beta_2),\ldots, \rho(\beta_n)$ describes a branching tree, in which any path from the root to one of the final twigs represents a particular history of the entire causal diamond, coarse-grained on the appropriate timescale.  These histories are ``minimally decoherent'' in the sense that the only degrees of freedom that are traced over are those that cannot be accessed even in principle.  In practice, an observer at the time $\beta$ may well already assign a definite outcome to an observation even though no particles correlated with the apparatus have yet crossed $B_-(\beta)$.   There is a negligible but nonzero probability of recoherence until the first particles cross the boundary; only then is coherence irreversibly lost.

Strictly speaking, the above analysis should be expanded to allow for the different gravitational backreaction of different branches.  The exact location of the boundary $B$ at the time $\beta$ depends on what happens at later times.   (This suggests that ultimately, it may be more natural to construct the decoherent causal diamonds from the top down, starting in the future and ending in the past.)  Here we will be interested mainly in the application to the eternally inflating multiverse,\footnote{However, the above discussion has implications for any global geometry in which observers fall out of causal contact at late times, including crunching universes and black hole interiors.  Suppose all observers were originally causally connected, i.e., their past light-cones substantially overlap at early times.  Then the different classes of decoherent histories that may be experienced by different observers arise from differences in the amount, the identity, and the order of the degrees of freedom that the observer must trace over.} where we can sidestep this issue by choosing large enough timesteps.  In de~Sitter vacua, on timescales of order $t_\Lambda \sim |\Lambda|^{-1/2}$, the apparent horizon, which is locally defined, approaches the event horizon $B$ at an exponential rate.  Mathematically, the difference between the two depends on the future evolution, but it is exponentially small and thus is irrelevant physically.  Vacua with negative cosmological constant crunch on the timescale $t_\Lambda$~\cite{CDL} and so will not be resolved in detail at this level of coarse-graining.

We expect that the distinction between causal diamond bulk and its boundary is precise only to order $e^{-A(\beta)}$, where $A$ is the area of the boundary at the time $\beta$.  Because of entropy bounds~\cite{Tho93,Sus95,CEB1,CEB2}, no observables in any finite spacetime region can be defined to better accuracy than this.  A related limitation applies to the objective notion of decoherence we have given, and it will be inherited by the reconstruction of global geometry we propose below.   This will play an imporant role in Sec.~\ref{sec-hat}, where we will argue that the hat regions with $\Lambda=0$ provide an exact counterpart to the approximate observables and approximate decoherence described here.

\subsection{Global-local measure duality}
\label{sec-dual}

In this section, we will review the duality that relates the causal diamond to a global time cutoff called {\em light-cone time}: both define the same probabilities when they are used as regulators for eternal inflation.  As originally derived, the duality assumed the standard global picture as a starting point, a viewpoint we have criticized in Sec.~\ref{sec-global}.   Here we will take the opposite viewpoint: the local picture is the starting point, and the duality suggests that a global spacetime can be reconstructed from the more fundamental structure of decoherent causal diamond histories.  Indeed, light-cone time will play a central role in the construction proposed in Sec.~\ref{sec-everett}.

By restricting to a causal diamond, we obtained a natural choice of environment: the degrees of freedom that exit from the diamond.  Tracing over this environment leads to a branching tree of objective, observer-independent decoherent histories---precisely the kind of notion that was lacking in the global description.  In the causal diamond, bubbles of different vacua really do nucleate at specific times and places.  They decohere when the bubble wall leaves the diamond.  

Consider a large landscape of vacua.  Starting, say, in a vacuum with very large cosmological constant, a typical diamond contains a sequence of bubble nucleations (perhaps hundreds in some toy models~\cite{BP,BouYan07}), which ends in a vacuum with negative cosmological constant (and thus a crunch), or with vanishing cosmological constant (a supersymmetric open universe, or ``hat'').   Different paths through the landscape are followed with probabilities determined by branching ratios.  Some of these paths will pass through long-lived vacua with anomalously small cosmological constant, such as ours.

The causal diamond has already received some attention in the context of the multiverse.  It was proposed~\cite{Bou06} as a probability measure: a method for cutting off infinities and obtaining well-defined amplitudes.  Phenomenologically, the causal diamond measure is among the most successful proposals extant~\cite{BouHar07,BouLei09,BouHal09,BouHar10,BouFre10d}.  From the viewpoint of economy, it is attractive since it merely exploits a restriction that was already imposed on us by black hole complementarity and uses it to solve another problem.  And conceptually, our contrasting analyses of decoherence in the global and causal diamond viewpoints suggests that the causal diamond is the more fundamental of the two.  

This argument is independent of black hole complementarity, though both point at the same conclusion.  It is also independent of the context of eternal inflation.  However, if we assume that the universe is eternally inflating, then it may be possible to merge all possible causal diamond histories into a single global geometry.

\begin{figure*}
\begin{center}
\includegraphics[scale = .6]{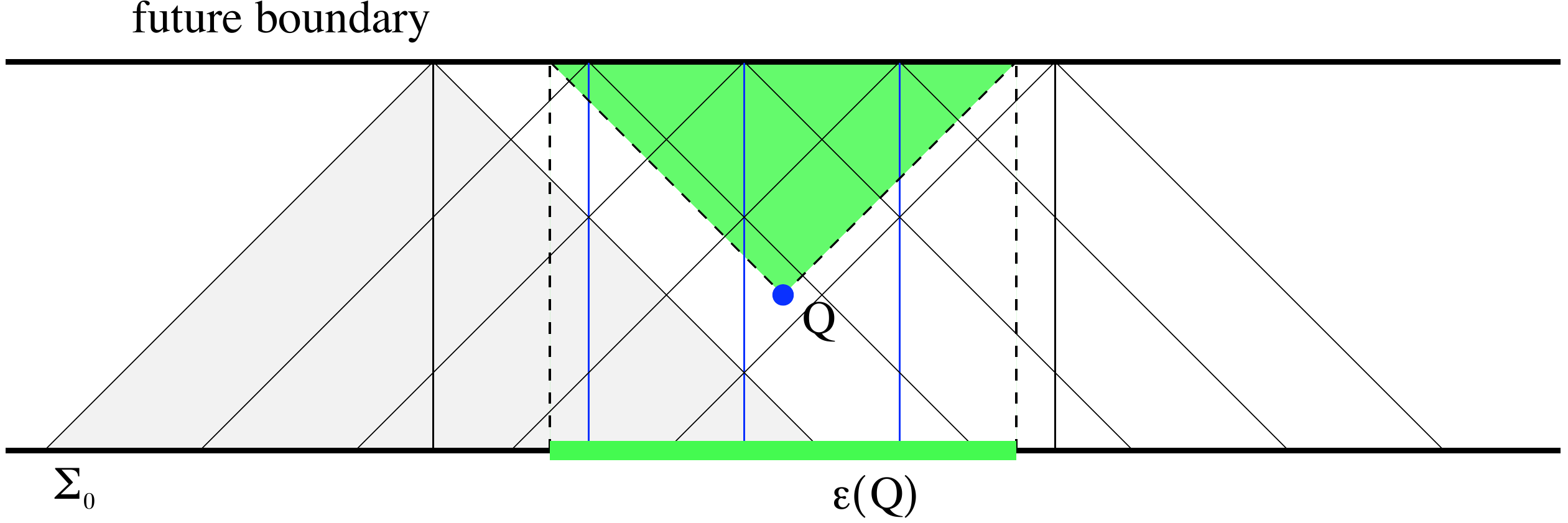}
\end{center}
\caption{In this diagram, the standard global picture of eternal inflation is taken as a starting point.  Geodesics (thin vertical lines) emanating from an initial surface $\Sigma_0$ define an ensemble of causal patches (the leftmost is shaded grey/light) with a particular mix of initial conditions. In the limit of a dense family of geodesics, the global spacetime is thus deconstructed into overlapping causal patches.  The number of patches overlapping at $Q$ is determined by the number of geodesics entering the future light-cone of $Q$.  In the continuum limit, this becomes the volume $\epsilon(Q)$ from which the relevant geodesics originate, which in turn defines the {\em light-cone time\/} at $Q$.  This relation implies an exact duality between the causal patch measure and the light-cone time cut-off for the purpose of regulating divergences and computing probabilities in eternal inflation.}
\label{fig-magic}
\end{figure*} 
If we are content to take the standard global picture as a starting point, then it is straightforward to deconstruct it into overlapping causal diamonds or patches\footnote{A causal patch can be viewed as the upper half of a causal diamond.  In practice the difference is negligible but strictly, the global-local duality holds for the patch, not the diamond.  Here we use it merely to motivate the construction of the following subsection, which does use diamonds.}~\cite{BouYan09,Nom11} (see Fig.~\ref{fig-magic}, taken from Ref.~\cite{BouYan09}). Indeed, a stronger statement is possible: as far as any prediction goes, the causal diamond viewpoint is indistinguishable from a particular time cutoff on the eternally inflating global spacetime.  An exact duality~\cite{BouYan09} dictates that relative probabilities computed from the causal diamond agree exactly with the probabilities computed from the light-cone time cutoff.\footnote{It is worth noting that the light-cone time cutoff was not constructed with this equivalence in mind.  Motivated by an analogy with the UV/IR relation of the AdS/CFT correspondence~\cite{GarVil08}, light-cone time was formulated as a measure proposal~\cite{Bou09} before the exact duality with the causal diamond was discovered~\cite{BouYan09}.}  The duality picks out particular initial conditions for the causal diamond: it holds only if one starts in the ``dominant vacuum'', which is the de~Sitter vacuum with the longest lifetime. 

The light-cone time of an event $Q$ is defined~\cite{Bou09} in terms of the volume $\epsilon(Q)$ of the future light-cone of $Q$ on the future conformal boundary of the global spacetime; see Fig.~\ref{fig-magic}:
\begin{equation}
t(Q)\equiv -\frac{1}{3} \log \epsilon(Q)~.
\end{equation}
The volume $\epsilon$, in turn, is defined as the proper volume occupied by those geodesics orthogonal to an initial hypersurface $\Sigma_0$ that eventually enter the future of $Q$.  (For an alternative definition directly in terms of an intrinsic boundary metric, see Ref.~\cite{BouFre10b}.)  We emphasize again that in these definitions, the standard global picture is taken for granted; we disregard for now the objections of Sec.~\ref{sec-global}.

The light-cone time of an event tells us the factor by which that event is overcounted in the overlapping ensemble of diamonds.  This follows from the simple fact that the geodesics whose causal diamonds includes $Q$ are precisely the ones that enter the causal future of $Q$.  Consider a discretization of the family of geodesics orthogonal to $\Sigma_0$ into a set of geodesics at constant, finite density, as shown in Fig.~\ref{fig-magic}.  The definition of light-cone time ensures that the number of diamonds that contain a given event $Q$ is proportional to $\epsilon=\exp[-3t(Q)]$.  Now we take the limit as the density of geodesics on $\Sigma_0$ tends to infinity.   In this limit, the entire global spacetime becomes covered by the causal diamonds spanned by the geodesics.  The relative overcounting of events at two different light-cone times is still given by a factor $\exp(-3\Delta t)$.  (To show that this implies the exact equivalence of the causal diamond and the light-conetime cutoff, one must also demonstrate that the rate at which events of any type $I$ occur depends only on $t$. This is indeed the case if $\Sigma_0$ is chosen sufficiently late, i.e., if initial conditions on the patches conform to the global attractor regime.   Since we are not primarily interested in the measure problem here, we will not review this aspect of the proof; see Ref.~\cite{BouYan09} for details.)   

Given the above deconstruction of the global spacetime, it is tempting to identify the eternally inflating multiverse with the many worlds of quantum mechanics, if the latter could somehow be related to branches in the wavefunction of the causal diamonds.  Without decoherence, however, there is neither a consistent global picture (as shown in Sec.~\ref{sec-global}) nor a sensible way of picking out a preferred basis that would associate ``many worlds'' to the pure quantum state of a causal diamond (Sec.~\ref{sec-patch}).\footnote{In this aspect our viewpoint differs from~\cite{Nom11}.}

We have already shown that decoherence at the causal diamond boundary leads to distinct causal diamond histories or ``worlds''.  To recover the global multiverse and demonstrate that it can be viewed as a representation of the many causal diamond worlds, one must show that it is possible to join together these histories consistently into a single spacetime.  This task is nontrivial.  In the following section we offer a solution in a very simple setting; we leave generalizations to more realistic models to future work.  Our construction will {\em not} be precisely the inverse of the deconstruction shown in Fig.~\ref{fig-magic}; for example, there will be no overlaps.  However, it is closely related; in particular, we will reconstruct the global spacetime in discrete steps of light-cone time.

\subsection{Constructing a global multiverse from many causal diamond worlds}
\label{sec-everett}

\begin{figure*}
\begin{center}
\includegraphics[scale = .35]{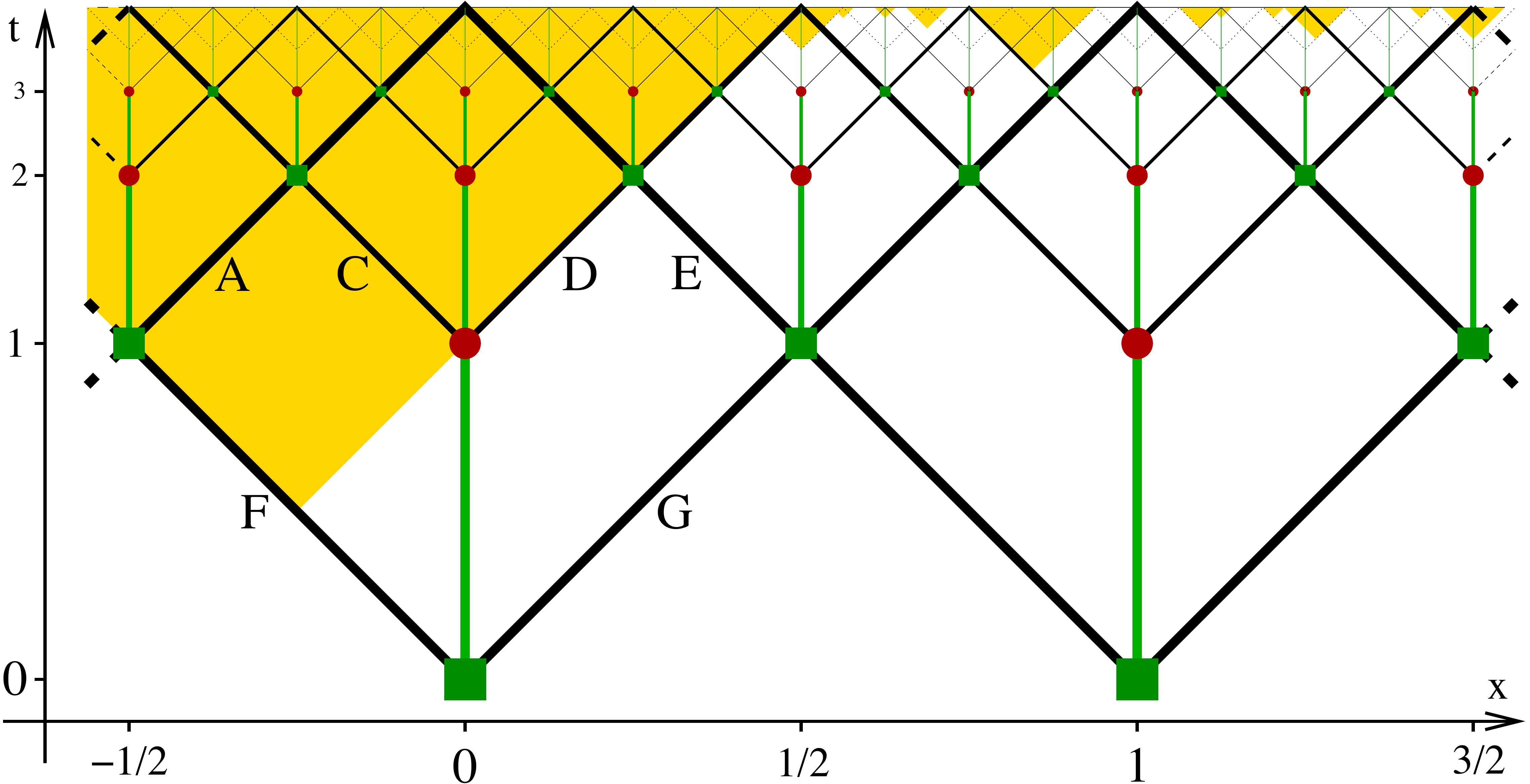}
\end{center}
\caption{Construction of a global spacetime from decoherent causal
  diamond histories.  Green squares indicate the origins of causal
  diamonds that tile the global de~Sitter space.  Red dots indicate
  discrete time steps of order $t_\Lambda$ at which decoherence occurs
  within individual causal diamonds.  For example, a nonzero amplitude
  for bubble nucleation in the region bounded by $ACDEFG$ entangles
  the null segment $A$ with $C$: either both will contain a bubble or
  neither will.  Similarly, $D$ is entangled with $E$.  The density
  matrix for $CD$ is diagonal in a basis consisting of pure states in
  which a bubble either forms or does not form.  This eliminates
  superpositions of the false (white) and true (yellow) vacuum.
  Initial conditions for new causal diamonds at green squares are
  controlled by entanglements such as that of $A$ with $C$.  Further
  details are described in the text.}
\label{fig-patches}
\end{figure*}

In this section, we will sketch a construction in $1+1$ dimensions by which a global picture emerges in constant increments of light-cone time (Fig.~\ref{fig-patches}).   For simplicity, we will work on a fixed de~Sitter background metric,
\begin{equation}
\frac{ds^2}{\ell^2} = -(\log 2~dt)^2+2^{2t}~dx^2~,
\label{eq-11m}
\end{equation}
where $\ell$ is an arbitrary length scale.  A future light-cone from an event at the time $t$ grows to comoving size $\epsilon = 2^{1-t}$, so $t$ represents light-cone time up to a trivial shift and rescaling: $t=1-\log_2 \epsilon$.  We take the spatial coordinate to be noncompact, though our construction would not change significantly if $x$ was compactified by identifying $x\cong x+n$ for some integer $n$.

The fixed background assumption allows us to separate the geometric problem---building the above global metric from individual causal diamonds---from the problem of matching matter configurations across the seams.  We will be interested in constructing a global spacetime of infinite four-volume in the future but not int the past.  Therefore, we take the geodesic generating each diamond to be maximally extended towards the future, but finite towards the past.  This means that the lower tips do {\em not\/} lie on the past conformal boundary of de~Sitter space.  Note that all such semi-infinite diamonds are isometric.

The geometric problem is particularly simple in 1+1 dimensions because it is possible to tile the global geometry precisely with causal diamonds, with no overlaps.  We will demonstrate this by listing explicitly the locations of the causal diamond tiles.  All diamonds in our construction are generated by geodesics that are comoving in the metric of Eq.~(\ref{eq-11m}): if the origin of the diamond is at $(t,x)$, then its tip is at $(\infty,x)$.  Hence we will label diamonds by the location of their origins, shown as green squares in Fig.~\ref{fig-patches}.  They are located at the events
\begin{eqnarray}
(x,t) & = & (m,0) \\
(x,t) & = & \left(\frac{2m+1}{2^n},n \right)~,
\end{eqnarray} 
where $n$ runs over the positive integers and $m$ runs over all integers.  From Fig.~\ref{fig-patches}, it is easy to see that these diamonds tile the global spacetime, covering every point with $t\geq 1$ precisely once, except on the edges.  We now turn to the task of matching the quantum states of matter at these seams.

We will assume that there exists a metastable vacuum $\cal{F}$ which decays with a rate $\Gamma$ per unit Hubble time and Hubble volume to a terminal vacuum $\cal{T}$ which does not decay.  We will work to order $\Gamma$, neglecting collisions between bubbles of the $\cal{T}$ vacuum.  A number of further simplifications are made below.

Our construction will be iterative.  We begin by considering the causal diamond with origin at $(x,t)=(0,0)$.  In the spirit of Sec.~\ref{sec-patch}, we follow the generating geodesic (green vertical line) for a time $\Delta t=1$ (corresponding to a proper time of order $t_\Lambda $) to the event $t=1$, $x=0$, marked by a red circle in Fig.~\ref{fig-patches}.  The future light-cone of this event divides the boundary of the $(0,0)$ diamond into two portions, $B_\pm$, as in Fig.~\ref{fig-beta}.  $B_-$ itself consists of two disconnected portions, which we label $A$ and $E$.  Together with segments $C$ and $D$ of the future light-cone, $ACDE$ forms a Cauchy surface of the diamond.  The evolution from the bottom boundary $FG$ to the surface $ACDE$ is unitary.  For definiteness, we will assume that the state on $FG$ is the false vacuum, though other initial conditions can easily be considered.

The pure state on $ACDE$ can be thought of as a superposition of the false vacuum with bubbles of true vacuum that nucleate somewhere in the region delimited by $ACDEFG$.  To keep things simple, we imagine that decays can occur only at three points, at each with probability $\bar\Gamma\sim \Gamma/3$: at the origin, $(0,0)$; at the midpoint of the edge $F$, $(-\frac{1}{4},\frac{1}{2})$, and at the midpoint of $G$, $(\frac{1}{4},\frac{1}{2})$.  We assume, moreover, that the true vacuum occupies the entire future light-cone of a nucleation point.  In this approximation the pure state on $ACDE$ takes the form
\begin{eqnarray}
  (1-3\bar\Gamma)^{1/2} & |{\cal F}\rangle_A |{\cal F}\rangle_C |{\cal F}\rangle_D |{\cal F}\rangle_E & + \bar \Gamma^{1/2} |{\cal T}\rangle_A |{\cal T}\rangle_C |{\cal F}\rangle_D |{\cal F}\rangle_E \nonumber \\
  +~ \bar\Gamma^{1/2} & |{\cal T}\rangle_A |{\cal T}\rangle_C |{\cal T}\rangle_D |{\cal T}\rangle_E & +\bar \Gamma^{1/2} |{\cal F}\rangle_A |{\cal F}\rangle_C |{\cal T}\rangle_D |{\cal T}\rangle_E~,  
\label{eq-ACDE}
\end{eqnarray}
where the last three terms correspond to the possible nucleation points, from left to right.

From the point of view of an observer in the $(0,0)$ diamond, the Hilbert space factor $AE$ should be traced out.  This results in a density matrix on the slice $CD$, which can be regarded as an initial condition for the smaller diamond beginning at the point $(x,t)=(0,1)$:
\begin{eqnarray} 
\rho(0,1) & =   (1-3\bar\Gamma) & |{\cal F}\rangle_C |{\cal F}\rangle_D \mbox{~}_D\langle {\cal F}| \mbox{\,}_C \langle {\cal F}| \nonumber \\
& +~\bar\Gamma  & |{\cal F}\rangle_C |{\cal T}\rangle_D \mbox{~}_D\langle {\cal T}| \mbox{\,}_C \langle {\cal F}| \nonumber \\
& +~\bar\Gamma & |{\cal T}\rangle_C |{\cal F}\rangle_D \mbox{~}_D\langle {\cal F}| \mbox{\,}_C \langle {\cal T}| \nonumber \\
& +~\bar\Gamma & |{\cal T}\rangle_C |{\cal T}\rangle_D \mbox{~}_D\langle {\cal T}| \mbox{\,}_C \langle {\cal T}|
\label{eq-CD}
\end{eqnarray}
The density matrix can be regarded as an ensemble of four pure states: $|{\cal F}\rangle_C |{\cal F}\rangle_D$ with probability $(1-3\bar\Gamma)$; and $|{\cal F}\rangle_C |{\cal T}\rangle_D, |{\cal T}\rangle_C |{\cal F}\rangle_D, |{\cal T}\rangle_C |{\cal T}\rangle_D $, each with probability $\bar\Gamma$.

The same construction can be applied to every ``zeroth generation'' causal diamond: the diamonds with origin at $(m,0)$, with $m$ integer.  Since their number is infinite, we can  realize the ensemble of Eq.~(\ref{eq-CD}) precisely, in the emerging global spacetime, by assigning appropriate initial conditions to the ``first generation sub-diamonds''  $(m,1)$.  The state  $|{\cal F}\rangle_C |{\cal T}\rangle_D$ is assigned to a fraction $1-3\bar\Gamma$ of the $(m,1)$ diamonds; and each of the states  $|{\cal F}\rangle_C |{\cal T}\rangle_D, |{\cal T}\rangle_C |{\cal F}\rangle_D, |{\cal T}\rangle_C |{\cal T}\rangle_D $  is assigned to a fraction $\bar\Gamma$ of $(m,1)$ diamonds.\footnote{We defer to future work the interesting question of whether further constraints should be imposed on the statistics of this distribution.  For example, for rational values of $\bar\Gamma$, the assignment of pure states to diamonds could be made in a periodic fashion, or at random subject only to the above constraint on relative fractions.}

So far, we have merely carried out the process described in Fig.~\ref{fig-beta} for one time step in each of the $(m,0)$ diamonds, resulting in initial conditions for the subdiamonds that start at the red circles at $(m,1)$.  In order to obtain a global description, we must also ``fill in the gaps'' between the $(m,1)$ diamonds by specifying initial conditions for the ``first generation new diamonds'' that start at the green squares at $(m+\frac{1}{2},1)$.  But their initial conditions are completely determined by the entangled pure state on $ACDE$, Eq.~(\ref{eq-ACDE}), and the identical pure states on the analogous Cauchy surfaces of the other $(m,0)$ diamonds.  Because of entanglement, the state on $C$ is the same as on $A$.  If $C$ is in the true vacuum, then so is $A$; and if $C$ is in the false vacuum, then so is $A$.  The edges $D$ and $E$ are similarly entangled.   Thus, the assignment of definite initial conditions to the $(m,1)$ diamonds completely determines the initial conditions on the $(m+\frac{1}{2})$ diamonds.  We have thus generated initial conditions for all first-generation diamonds (those with $n=1$).  Now we simply repeat the entire procedure to obtain initial conditions for the second generation ($n=2$), and so on.\footnote{Note that this construction determines the initial conditions for all but a finite number of diamonds.  In a more general setting, it would select initial conditions in the dominant vacuum.  We thank Ben Freivogel and I-Sheng Yang for pointing out a closely related observation.}

This will generate a fractal distribution of true vacuum bubbles, of the type that is usually associated with a global description (Fig.~\ref{fig-patches}).  The manner in which this picture arises is totally distinct from a naive unitary evolution of global time slices, in which a full superposition of false and true vacuum would persist (with time-dependent amplitudes).  The standard global picture can only be obtained by exploiting the decoherence of causal diamonds while proliferating their number.  The multiverse is a patchwork of infinitely many branching histories, the many worlds of causal diamonds.

The construction we have given is only a toy model.   The causal diamonds in higher dimensions do not fit together neatly to fill the space-time as they do in $1+1$ dimensions, so overlaps will have to be taken into account.\footnote{Although not in the context of the multiverse, Banks and Fischler have suggested that an interlocking collection of causal diamonds with finite dimensional Hilbert spaces can be assembled into the global structure of de Sitter space~\cite{BanFis01b}.}  Moreover, we have not considered the backreaction on the gravitational field.  Finally, in general the future boundary is not everywhere spacelike but contains hats corresponding to supersymmetric vacua with $\Lambda=0$.  Our view will be that the hat regions play a very special role that is complementary, both in the colloquial and in the technical sense, to the construction we have given here.  Any construction involving finite causal diamonds in necessarily approximate.  We now turn to the potentially precise world of the Census Taker, a fictitious observer living in a hat region.

\section{The many worlds of the census taker}
\label{sec-hat}

\subsection{Decoherence and recoherence}
\label{sec-reverse}

In Sec.~\ref{sec-future}, we noted that decoherence is subjective to the extent that the choice of environment is based merely on the practical considerations of an actual observer.  We then argued in Sec.~\ref{sec-patch} that the boundary of a causal patch can be regarded as a preferred environment that leads to a more objective form of decoherence.  However, there is another element of subjectivity in decoherence which we have not yet addressed: decoherence is reversible.  Whether, and how soon, coherence is restored depends on the dynamical evolution governing the system and environment.

\begin{figure*}
\begin{center}
\includegraphics[scale = .35]{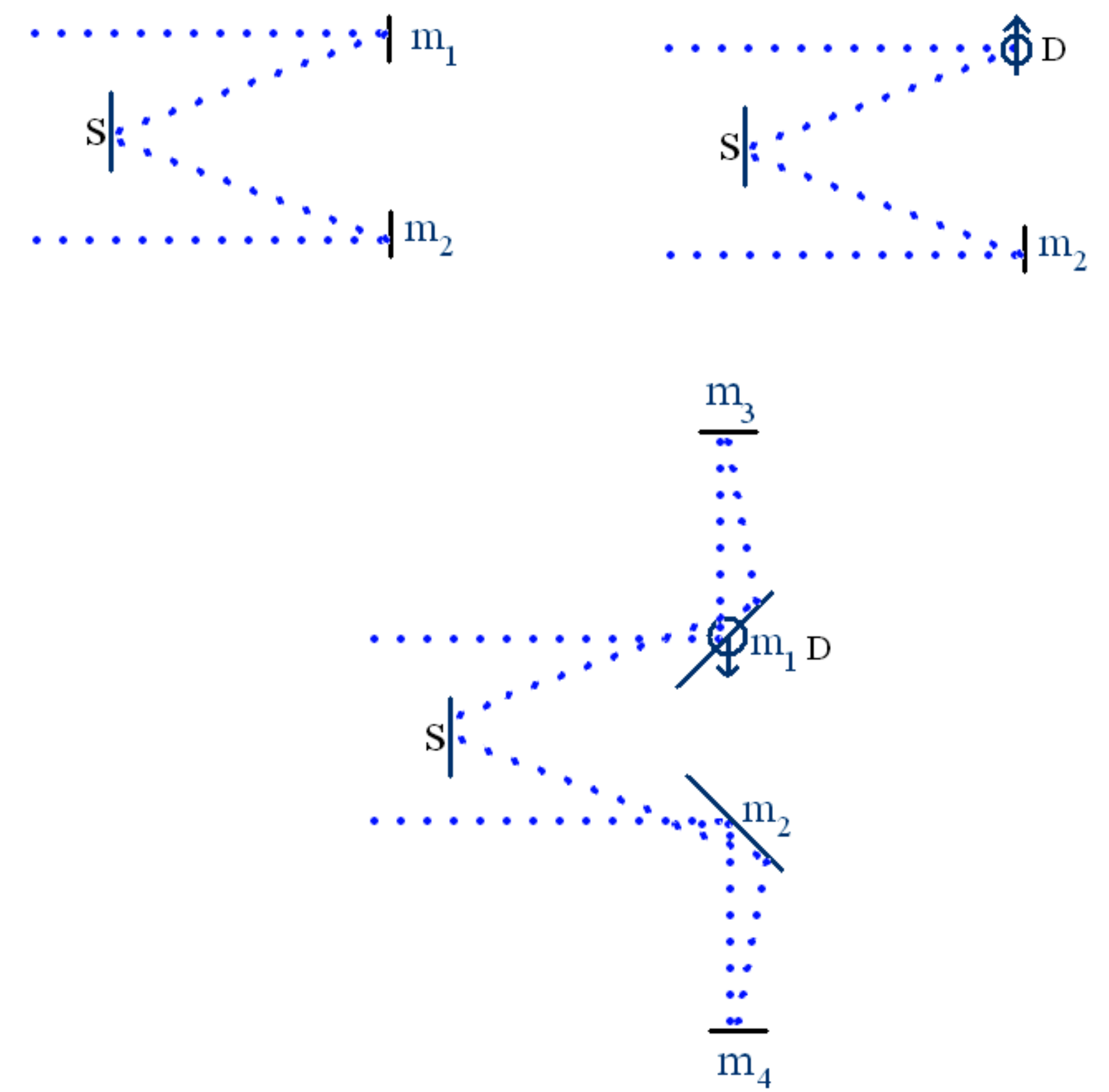}
\end{center}
\caption{Three optical interference experiments: $m$'s denote mirrors, $S$ is a screen, and $D$ means detector.  In the first setup, the coherence of the two paths is maintained.  In the second setup, a detector is placed at one mirror.  If the detector is treated as an environment and its Hilbert space is traced over, then the coherence of the superposition of photon paths is lost and the interference pattern disappears.   In the third setup, the detector reverts to its original state when the photon passes it a second time.   During the (arbitrarily long) time when the photon travels between $m_1D$ and $m_3$, tracing over the detector leads to decoherence.  But after the second pass through $m_1D$, coherence is restored, so the screen will show the same pattern as in the first setup.}
\label{fig-mirrors}
\end{figure*}

Consider an optical interference experiment (shown in Fig.~\ref{fig-mirrors}) in which a light beam reflects off two mirrors, $m_1$ and $m_2$, and then illuminates a screen $S$. There is no sense in which one can say that a given photon takes one or the other routes.

On the other hand if an observer or simple detector $D$ interacts with the photon and records which mirror the photon bounced off, the interference is destroyed. One of the two possibilities is made real by the interaction and thereafter the observer may ignore the branch of the wave function which does not agree with the observation.  Moreover, if a second observer describes the whole experiment ( including the first observer) as a single quantum system, the second observer's observation will entangle it in a consistent way.

Now let's consider an unusual version of the experiment in which the upper arm of the interferometer is replaced by a mirror-detector $m_1D$ which detects a photon and deflects it toward mirror $m_3.$ From $m_3$ the photon is reflected back to the detector and then to the screen. The detector is constructed so that if a single photon passes through it, it flips to a new state (it detects) but the next time a photon passes through, it flips back to the original state.  The detector is well-insulated from any environment.  The lower arm of the interferometer also has the path length increased but without a detector.

Since if the photon goes through the detector, it goes through twice, at the end of the experiment the detector is left in the original state of no-detection.  It is obvious that in this case the interference is restored.  But there is something unusual going on. During an intermediate time interval the photon was entangled with the detector. The event of passing through the upper arm has been recorded and the photon's wave function has collapsed to an incoherent superposition. But eventually the photon and the detector are disentangled. What happened was made to unhappen.

This illustrates that in order to give objective meaning to an event such as the detection of a photon, it is not enough that the system becomes entangled with the environment: the system must become {\em irreversibly} entangled with the environment, or more precisely, with {\em some} environment.

\subsection{Failure to irreversibly decohere: A limitation of finite systems}
\label{sec-limitation}

The above example may seem contrived, since it relied on the perfect isolation of the detector from any larger environment, and on the mirror $m_3$ that ensured that the detected photon cannot escape.  It would be impractical to arrange in a similar manner for the recoherence of macroscopic superpositions, since an enormous number of particles would have to be carefully controlled by mirrors.  However, if we are willing to wait, then recoherence is actually inevitable in any system that is {\em dynamically closed}, i.e., a system with finite maximum entropy at all times.

For example consider a world inside a finite box, with finite energy and perfectly reflecting walls. If the box is big enough and is initially out of thermal equilibrium, then during the return to equilibrium structures can form, including galaxies, planets, and observers. Entanglements can form between subsystems, but it is not hard to see that they cannot be irreversible. Such closed systems with finite energy have a finite maximum entropy, and for that reason the state vector will undergo quantum recurrences. Whatever state the system
finds itself in, after a suitably long time it will return to the same state or to an arbitrarily close state.  The recurrence time is bounded by $\exp(N)=\exp(e^{S_{\rm max}})$, where $N$ is the dimension of the Hilbert space that is explored and $S_{\rm max}$ is the maximum entropy.

This has an important implication for the causal diamonds of Sec.~\ref{sec-patch}.   We argued that the diamond bulk decoheres when degrees of freedom cross the boundary $B$ of the diamond.   But consider a potential that just contains a single, stable de~Sitter vacuum with cosmological constant $\Lambda$.   Then the maximum area of the boundary of the diamond is the horizon area of empty de~Sitter space, and the maximum entropy is $S_{\rm max}=A_{\rm max}/4=3\pi/\Lambda$.   This is the maximum {\em total\/} entropy~\cite{Bou00a}, which is the sum of the matter entropy in the bulk and the Bekenstein-Hawking entropy of the boundary.  Assuming unitarity and ergodicity~\cite{DysKle02}, this system is dynamically closed, and periodic recurrences are inevitable.  (See Ref.~\cite{BanFis01a,BanFis02,BanFis04} for a discussion of precision in de Sitter space.)

Next, consider a landscape that contains vacua with positive and negative cosmological constant.  We assume, for now, that there are no vacua with $\Lambda=0$.  Then the maximum area $A_{\rm max}$ of the causal diamond boundary $B$ is given by the greater of $\Lambda_+^{-1}$ and $\Lambda_-^{-2}$, where $\Lambda_+$ is the smallest positive value of $\Lambda$ among all landscape vacua, and $\Lambda_-$ is the largest negative value~\cite{BouFre10a}.   $B$ is a null hypersurface with two-dimensional spatial cross-sections, and it can be thought of as the union of two light-sheets that emanate from the cross-section of maximum area.  Therefore, the entropy that passes through $B$ is bounded by $A_{\rm max}/2$ and hence is finite.

What does finite entropy imply for the decoherence mechanism of Sec.~\ref{sec-patch}?  If the causal diamond were a unitary, ergodic quantum system, then it would again follow that recurrences, including the restoration of coherence, are inevitable.  This is plausible for causal diamonds that remain forever in inflating vacua, but such diamonds form a set of measure zero.  Generically, causal diamonds will end up in terminal vacua with negative cosmological constant, hitting a big crunch singularity after finite time.  In this case it is not clear that they admit a unitary quantum mechanical description over arbitrarily long timescales, so recurrences are not mandatory.  However, since ergodicity cannot be assumed in this case, it seems plausible to us that there exists a tiny non-zero probability for recurrences.  We expect that in each metastable de~Sitter vacuum, the probability for re-coherence is given by the ratio of the decay timescale to the recurrence timescale.  Typically, this ratio is super-exponentially small~\cite{BouFre06b}, but it does not vanish.  In this sense, the objective decoherence of causal diamond histories described in Sec.~\ref{sec-patch} is not completely sharp.

\subsection{Sagredo's postulates}
\label{sec-sagredo}

{\em The next morning, Simplicio and Salviati visit Sagredo to continue their discussion.}

{\sc Simplicio:} I have been pondering the idea we came up with yesterday, and I am convinced that we have completely solved the problem.  Causal diamonds have definite histories, obtained by tracing over their boundary, which we treat as an observer-independent environment.  This gets rid of superpositions of different macroscopic objects, such as bubbles of different vacua, without the need to appeal to actual observers inside the diamond.  Each causal diamond history corresponds to a sequence of things that ``happen''.  And the global picture of the multiverse is just a representation of all the possible diamond histories in a single geometry: the many worlds of causal diamonds!

{\sc Sagredo:} I wish I could share in your satisfaction, but I am uncomfortable.  Let me describe my concerns, and perhaps you will be able to address them.

{\sc Salviati:}  I, too, have been unable to shake off a sense that that this is not the whole story---that we should do better. I would be most interested to hear your thoughts, Sagredo.

{\sc Sagredo:}  It's true, as Simplicio says, that things ``happen'' when we trace over degrees of freedom that leave the causal diamond.  Pure states become a density matrix, or to put it in Bohr's language, the wavefunction describing the interior of the diamond collapses.  But how do we know that the coherence will not be restored?  What prevents things from ``unhappening'' later on?

{\sc Simplicio:} According to Bohr, the irreversible character of observation is due to the large classical nature of the apparatus.

{\sc Salviati:} And decoherence allows us to understand this rather vague statement more precisely: the apparatus becomes entangled with an enormous environment, which is infinite for all practical purposes.

{\sc Sagredo:} But even a large apparatus is a quantum system, and in principle, the entanglement can be undone. The irreversibility of decoherence is often conflated with the irreversibility of thermodynamics. A large system of many degrees of freedom is very unlikely to find its way back to a re-cohered state. However, thermodynamic irreversibility is an idealization that is only true for infinite systems. The irreversibility of decoherence, too, is an approximation that becomes exact only for infinite systems. 

{\sc Simplicio:} But think how exquisite the approximation is!  In a causal diamond containing our own history, the boundary area becomes as large as billions of light years, squared, or $10^{123}$ in fundamental units.  As you know, I have studied all the ancients.  I learned that the maximum area along the boundary of a past light-cone provides a bound on the size $N$ of the Hilbert space describing everything within the light-cone: $N\sim \exp(10^{123})$~\cite{Bou00a}.  And elsewhere I found that re-coherence occurs on a timescale $N^N\sim \exp[\exp(10^{123})]$~\cite{Zur03}.  This is much longer than the time it will take for our vacuum to decay~\cite{KKLT,FreLip08} and the world to end in a crunch.  So why worry about it?

{\sc Sagredo:}  It's true that re-coherence is overwhelmingly unlikely in a causal diamond as large as ours.  But nothing you said convinces me that the probability for things to ``unhappen'' is exactly zero.

{\sc Salviati:} To me, it is very important to be able to say that some things really do happen, irreversibly and without any uncertainty.  If this were not true, then how could we ever make sense of predictions of a physical theory?  If we cannot be sure that something happened, how can we ever test the prediction that it should or should not happen?

{\sc Sagredo:} That's it---this is what bothered me.  The notion that things really happen should be a fundamental principle, and the implementation of fundamental principles in a physical theory should not rely solely on approximations.  So let me define this more carefully in terms of a definition and a postulate.

\paragraph{Definition I}  Consider an instance of decoherence (or ``collapse of a wave function'') in a Hilbert space ${\cal H}_S$, which occurs as a result of entanglement with another Hilbert space ${\cal H}_E$.  The event will be said to {\em happen} if the entanglement between ${\cal H}_E$ and ${\cal H}_S$ is irreversible; and the system $S$ can then be treated as if it was in one of the pure states that constitute the basis that diagonalizes the density matrix obtained by tracing over the Hilbert space ${\cal H}_E$.

\paragraph{Postulate I} Things {\em happen}.\\

\noindent In other words, there exist some entanglements in Nature that will not be reversed with any finite probability.

{\sc Simplicio:}  Let me see if I can find an example that satisfies your postulate.  Suppose that an apparatus is in continuous interaction with some environment.  Even an interaction with an single environmental photon can record the event.  If the photon disappears to infinity so that no mirror can ever reflect it back, then the event has {\em happened}.

{\sc Sagredo:}  Your example makes it sound like irreversible decoherence is easy, but I don't think this is true.  For example, in the finite causal diamonds we considered, there is no ``infinity'' and so nothing can get to it.

{\sc Salviati:}  Sagredo's postulate is in fact surprisingly strong!  Any dynamically closed system (a system with finite entropy) cannot satisfy the postulate, because recurrences are inevitable.  This not a trivial point, since it immediately rules out certain cosmologies.   Stable de Sitter space is a closed system.  More generally, if we consider a landscape with only positive energy local minima, the recurrence time is controlled by the minimum with the smallest cosmological constant. So the recurrence time is finite, and nothing {\em happens}.  Anti de Sitter space is no better. As is well known, global AdS is a box with reflecting walls. At any finite energy it has finite maximum entropy and also gives rise to recurrences.

{\sc Simplicio:} I can think of a cosmology that satisfies Postulate I, along the lines of my previous example.  I will call it ``S-matrix cosmology''.  It takes place in an asymptotically flat spacetime and can described as a scattering event. The initial state is a large number of incoming stable particles. The particles could be atoms of hydrogen, oxygen, carbon, etc. The atoms come together and form a gas cloud that contracts due to gravity. Eventually it forms a solar system that may have observers doing experiments.  Photons scatter or are emitted from the apparatuses and become irreversibly entangled as they propagate to infinity in the final state. The central star collapses to a black hole and evaporates into Hawking radiation. Eventually everything becomes outgoing stable particles.

{\sc Sagredo:}  It is true that there are some things that {\em happen} in your S-matrix cosmology.  But I am not satisfied.  I think there is a larger point that Salviati made: we would like a cosmology in which it is possible to give a precise operational meaning to quantum mechanical predictions.  The notion that things unambiguously {\em happen} is necessary for this, but I now realize that it is not sufficient.

{\sc Salviati:} Now, I'm afraid, you have us puzzled.  What is wrong with the S-matrix cosmology?

{\sc Sagredo:}  Quantum mechanics makes probabilistic predictions; when something {\em happens}, each possible outcome has probability given by the corresponding diagonal entry in the density matrix.  But how do we verify that this outcome really happens with predicted probability?

{\sc Salviati:}  Probabilities are frequencies; they can be measured only by performing an experiment many times.  For example, to test the assertion that the probability for ``heads'' is one half, you flip a coin a large number of times and see if, within the margin of error, it comes up heads half of the time.  If for some reason it were only possible to flip a coin once there would be no way to test the assertion reliably.  And to be completely certain of the probability distribution, it would be necessary to flip the coin infinitely many times.

{\sc Simplicio:} Well, my S-matrix cosmology can be quite large.  For example, it might contain a planet on which someone flips a quantum coin a trillion times.    Photons record this information and travel to infinity.  A trillion outcomes {\em happen}, and you can do your statistics.  Are you happy now?

{\sc Sagredo:} A trillion is a good enough approximation to infinity for all practical purposes.  But as I said before, the operational testability of quantum-mechanical predictions should be a fundamental principle.  And the implementation of a fundamental principle should not depend on approximations.

{\sc Salviati:}  I agree.  No matter how large and long-lived Simplicio makes his S-matrix cosmology, there will only be a finite number of coin flips.  And the cosmology contains many ``larger'' experiments that are repeated even fewer times, like the explosion of stars.  So the situation is not much better than in real observational cosmology. For example, inflation tells us that the quadrupole anisotropy of the CMB has a gaussian probability distribution with a variance of a few times $10^{-5}$. But it can only be measured once, so we are very far from being able to confirm this prediction with complete precision.  

{\sc Sagredo:} Let me try to state this more precisely.   Quantum mechanics makes probabilistic predictions, which have the following operational definition:

\paragraph{Definition II}   Let $P(i)$ be the theoretical probability that outcome $i$ {\em happens} (i.e., $i$ arises as a result of irreversible decoherence), given by a diagonal entry in the density matrix.   Let $N$ be the number of times the corresponding experiment is repeated, and let $N_i$ be the number of times the outcome $i$ {\em happens}.  The sharp prediction of quantum mechanics is that 
\begin{equation}
P(i)=\lim_{N\to \infty}{N_i \over N}~.
\label{eq-nin}
\end{equation}
\mbox{}\\[-3ex]

{\sc Salviati:}  Do you see the problem now, Simplicio?  What bothers us about your S-matrix cosmology is that $N$ is finite for any experiment, so it is impossible to verify quantum-mechanical predictions with arbitrary precision.

{\sc Simplicio:}  Why not proliferate the S-matrix cosmology?  Instead of just one asymptotically flat spacetime, I will give you infinitely many replicas with identical in-states.   They are completely disconnected from one another; their only purpose is to allow you to take $N\to \infty$.

{\sc Salviati:} Before we get to the problems, let me say that there is one thing I like about this proposal: it provides a well-defined setting in which the many worlds interpretation of quantum mechanics appears naturally.  For example, suppose we measure the $z$-component of the spin of an electron.  I have sometimes heard it said that when decoherence occurs, the world into two equally real branches.  But there are problems with taking this viewpoint literally.  For example, one might conclude that the probabilities for the two branches are necessarily equal, when we know that in general they are not.

{\sc Simplicio:} Yes, I have always found this confusing.  So how does my proposal help?

{\sc Salviati:} The point is that in your setup, there are an infinite number of worlds to start with, all with the same initial conditions.  Each world within this collection does not split; the collection $S$ itself splits into two subsets.  In a fraction $p_i$ of worlds, the outcome $i$ happens when a spin is measured.  There is no reason to add another layer of replication.

{\sc Simplicio:} Are you saying that you reject the many worlds interpretation?

{\sc Salviati:} That depends on what you mean by it.  Some say that the many worlds interpretation is a theory in which reality is the many-branched wavefunction itself. I dislike this idea, because quantum mechanics is about observables like position,  momentum, or spin.  The wavefunction is merely an auxiliary quantity that tells you how to compute the probability for an actual observation.   The wavefunction itself is not a measurable thing. For example the wavefunction $\psi(x)$ of a particle cannot be measured, so in particular one cannot measure that it has split.  But suppose we had a system composed of an infinite number of particles, all prepared in an identical manner. Then the single particle wavefunction $\psi(x)$ becomes an observable for the larger system. For example, the single particle probability density $\psi^{\ast}(x)\psi(x)$ is a many-particle observable:
\begin{equation} 
\psi^{\ast}(x)\psi(x) = \lim_{N\to \infty} {1 \over N} \sum_{i=1}^N \delta(x_i -x)
\label{eq-multiop}
\end{equation} 

{\sc Simplicio:} I see.  Now if an identical measurement is performed on each particle, each single particle wavefunction splits, and this split wavefunction can be measured in the many-particle system.

{\sc Salviati:} To make this explicit, we could make the individual systems more complicated by adding a detector $d$ for each particle. Each particle-detector system can be started in the product state $\psi_0(x) \chi_0(d)$.  Allowing the particle to interact with the detector would create entanglement, i.e., a wavefunction of the form $\psi_1(x) \chi_1(d)+\psi_2(x) \chi_2(d)$.  But the branched wave function cannot be measured any better than the original one.  Now consider an unbounded number of particle-detector pairs, all starting in the same product state, and all winding up in entangled states. It is easy to construct operators analogous to Eq.~(\ref{eq-multiop}) in the product system that correspond to the single system's wave function.  So you see, that in the improved S-matrix cosmology there is no reason to add another layer of many-worlds.

{\sc Sagredo:} That's all very nice, but Simplicio's proposal does not help with making quantum mechanical predictions operationally well-defined.  You talk about many disconnected worlds, so by definition it is impossible to collect the $N\to \infty$ results and compare their statistics to the theoretical prediction.

{\sc Simplicio:} I see.  By saying that quantum mechanical predictions should be operationally meaningful, you mean not only that infinitely many outcomes {\em happen}, but that they are all accessible to an observer in a single universe.

{\sc Sagredo:} Yes, it seems to me that this requirement follows from Salviati's fundamental principle that predictions should have precise operational meaning.  Let me enshrine this in another postulate:

\paragraph{Postulate II}  Observables are {\em observable}.\\

\noindent By observables, I mean any Hermitian operators whose probability distribution is precisely predicted by a quantum mechanical theory from given initial conditions.  And by {\em observable}, I mean that the world is big enough that the observable can be measured infinitely many times by irreversible entanglement.

{\sc Salviati:} Like your first postulate, this one seems quite restrictive.  It obviously rules out the infinite set of S-matrix cosmologies, since products of field operators in different replicas are predicted but cannot be measured by any observers in any one cosmology.  And it gives us additional reasons to reject Simplicio's earlier suggestion: the single S-matrix cosmology contains observables such as the total number of outgoing particles, which cannot even {\em happen}, since there is no environment to measure them.

{\sc Simplicio:} Well, we were quite happy yesterday with the progress we had made on making decoherence objective in the causal diamond.  But these postulates of yours clearly cannot be rigorously satisfied in any causal diamond, no matter how large.  Perhaps it's time to compromise?  Are you really sure that fundamental principles have to have a completely sharp implementation in physical theory?  What if your postulates simply cannot be satisfied?

{\sc Salviati:} Yesterday, we did not pay much attention to causal diamonds that end in regions with vanishing cosmological constant.  Perhaps this was a mistake?  In such regions, the boundary area and entropy are both infinite.   I may not have not thought about it hard enough, but I see no reason why our postulates could not be satisfied in these ``hat'' regions. 

{\sc Sagredo:} Even if they can, there would still be the question of whether this helps make sense of the finite causal diamonds.  I care about this, because I think we live in one of them.

{\sc Salviati:} I understand that, but let us begin by asking whether our postulates might be satisfied in the hat.

\subsection{Irreversible decoherence and infinite repetition in the hat}
\label{sec-hats}

In the eternally inflating multiverse there are three types of causal diamonds. The first constitute a set of measure zero and remain forever in inflating regions. The entropy bound for such diamonds is the same as for the lowest-$\Lambda$ de Sitter space they access, and hence is finite.  The second type are the diamonds who end on singular crunches. If the crunches originated from the decay of a de Sitter vacuum then the entropy bound is again finite~\cite{HarSus10}.  Finally there are causal diamonds that end up in a supersymmetric bubble of zero cosmological constant. A timelike geodesic that enters such a bubble will remain within the diamond for an infinite time.  It is convenient to associate an observer with one of these geodesics, called the Census Taker.  

\begin{figure*}
\begin{center}
\includegraphics[scale = .45]{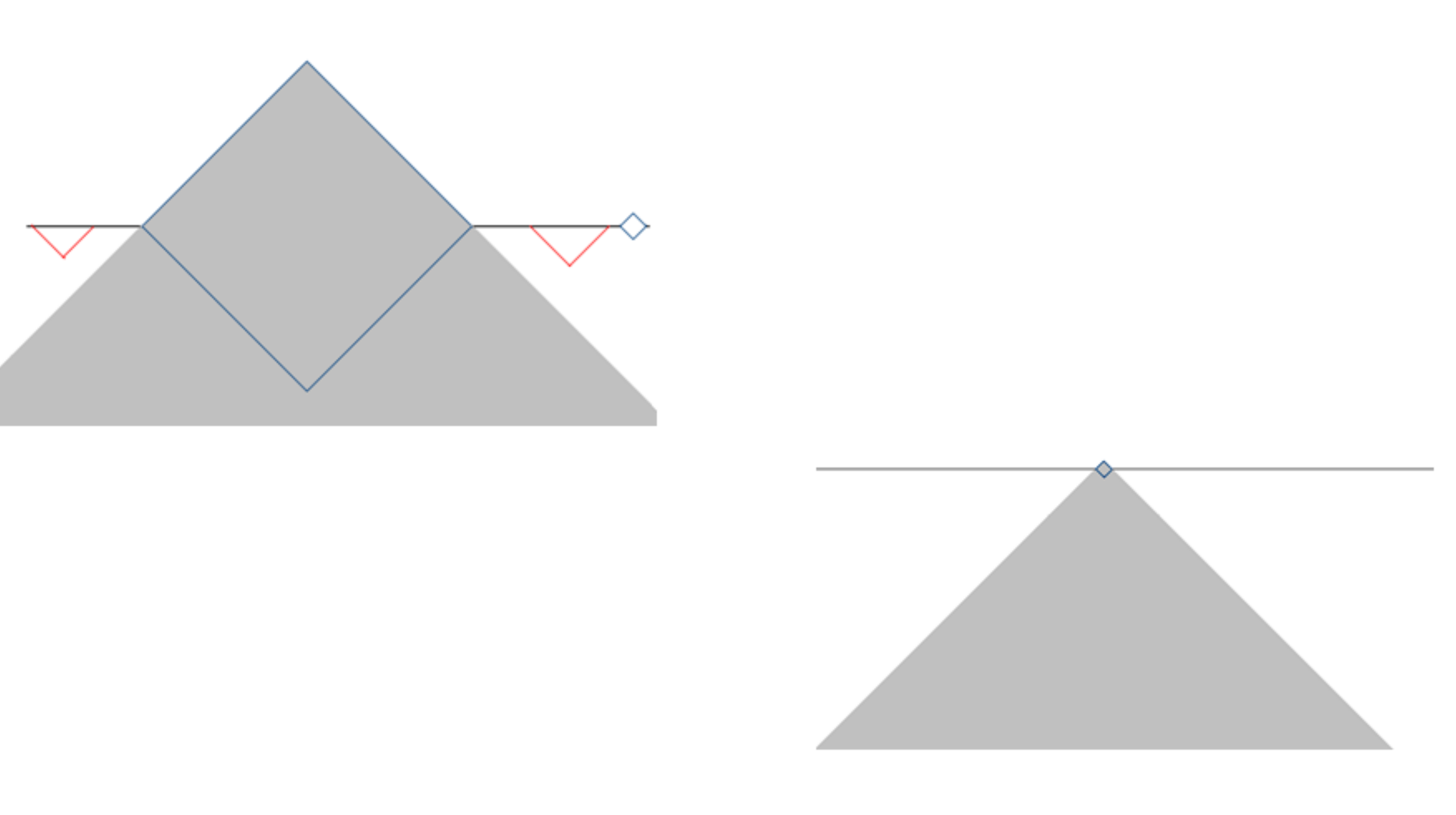}
\end{center}
\caption{The shaded region is the late time portion of a causal diamond that ends in a supersymmetric vacuum with vanishing cosmological constant (left).  We will think of this region as the asymptotic causal past of an ``observer'' in this vacuum, called the Census Taker.  A portion of the boundary of such causal diamonds has infinite cross-sectional area.  This portion, called a ``hat'', coincides with a light-like portion of the future conformal boundary in the standard global picture of the multiverse.  $\Lambda=0$ bubbles that form late look small on the conformal diagram (right).  But they always have infinite area.   Hence the hat regions can admit exact observables and can allow things to {\em happen}, with enough room for arbitrarily good statistics.}
\label{fig-biglittle}
\end{figure*}

The Census Taker's past light-cone becomes arbitrarily large at late times.  It asymptotes to the null portion of future infinity called the ``hat'', and from there extends down into the non-supersymmetric bulk of the multiverse (Fig.~\ref{fig-biglittle}).
If the CT's bubble nucleates very late, the entire hat appears very small on the conformal diagram and the CT's past light-cone looks similar to the past light-cone of a point on the future boundary of ordinary de Sitter space. However, this is misleading; the entropy bound for the hat geometry is not the de Sitter entropy of the ancestor de Sitter space. It is infinite.  

According to the criterion of Ref.~\cite{HarSus10}, the existence of light-sheets with unbounded area implies the existence of precise observables.  A holographic dual should exist by which these observables are defined.   The conjecture of Ref.~\cite{FreSek06}---the FRW/CFT duality---exhibits close similarities to AdS/CFT. In particular, the spatial hypersurfaces of the FRW geometry under the hat are open and thus have the form of three-dimensional Euclidean anti-de Sitter space.  Their shared conformal boundary---the space-like infinity of the FRW geometry---is the two-dimensional ``rim'' of the hat.   As time progresses, the past light-cone of the Census Taker intersects any fixed earlier time slice closer and closer to the boundary. For this reason the evolution of the Census Taker's observations has the form of a renormalization group flow in a two-dimensional Euclidean conformal field theory~\cite{SekSus09}.

We will return to the holographic dual later, but for now we will simply adopt the conjecture that exact versions of observables exist in the hat.  We can then ask whether hat observables are {\em observable}, in the sense of Postulate II above.  In particular, this would require that things {\em happen} in the hat, in the sense of Postulate I.  We do not understand the fundamental description of the hat well enough to prove anything rigorously, but we will give some arguments that make it plausible that both postulates can indeed be satisfied.  Our arguments will be based on conventional (bulk) observables.

Consider an interaction at some event $X$ in the hat region that entangles an apparatus with a photon.  If the photon gets out to the null conformal boundary, then a measurement has {\em happened} at $X$, and postulate I is satisfied.  In general the photon will interact with other matter, but unless this interaction takes the form of carefully arranged mirrors, it will not lead to recoherence.  Instead will enlarge the environment, and sooner or later a massless particle that is entangled with the event at $X$ will reach the null boundary.  For postulate I to be satisfied it suffices that there exist {\em some} events for which this is the case; this seems like a rather weak assumption.

By the FRW symmetry~\cite{CDL} of the hat geometry, the same measurement {\em happens}, with the same initial conditions, at an infinite number of other events $X_i$.  Moreover, any event in the hat region eventually enters the Census Taker's past light-cone.   Let $N(t)$ be the number of equivalent measurements that are in the Census Taker's past light-cone at the time $t$ along his worldline, and let $N_i(t)$ be the number of times the outcome $i$ {\em happens} in the same region.  Then the limit in Eq.~(\ref{eq-nin}) can be realized as
\begin{equation}
p_i= \lim_{t\to \infty} \frac{N_i(t)}{N(t)}~,
\end{equation}
and Postulate II is satisfied for the particular observable measured.  

A crucial difference to the S-matrix cosmology discussed in the previous subsection is that the above argument applies to any observable: if it happens once, it will happen infinitely many times.  It does not matter how ``large'' the systems are that participate in the interaction.  Because the total number of particles in the hat is infinite, there is no analogue of observables such as the total number of outgoing particles, which would be predicted by the theory but could not be measured.  This holds as long as the fundamental theory predicts directly only observables in the hat, which we assume.  Thus, we conclude that Postulate II is satisfied for all observables in the hat.

Since both postulates are satisfied, quantum mechanical predictions can be operationally verified by the Census Taker to infinite precision.  But how do the hat observables relate to the approximate observables in causal diamonds that do {\em not} enter hats, and thus to the constructions of Sec.~\ref{sec-nonhat}?  We will now argue that the non-hat observables are approximations that have exact counterparts in the hat.

\subsection{Black hole complementarity and hat complementarity}
\label{sec-complementarity}

In this subsection, we will propose a complementarity principle that relates exact Census Taker observables defined in the hat to the approximate observables that can be defined in other types of causal diamonds, which end in a crunch and have finite maximal area.  To motivate this proposal, we will first discuss black hole complementarity ~\cite{SusTho93} in some detail.

\begin{figure*}
\begin{center}
\includegraphics[scale = .45]{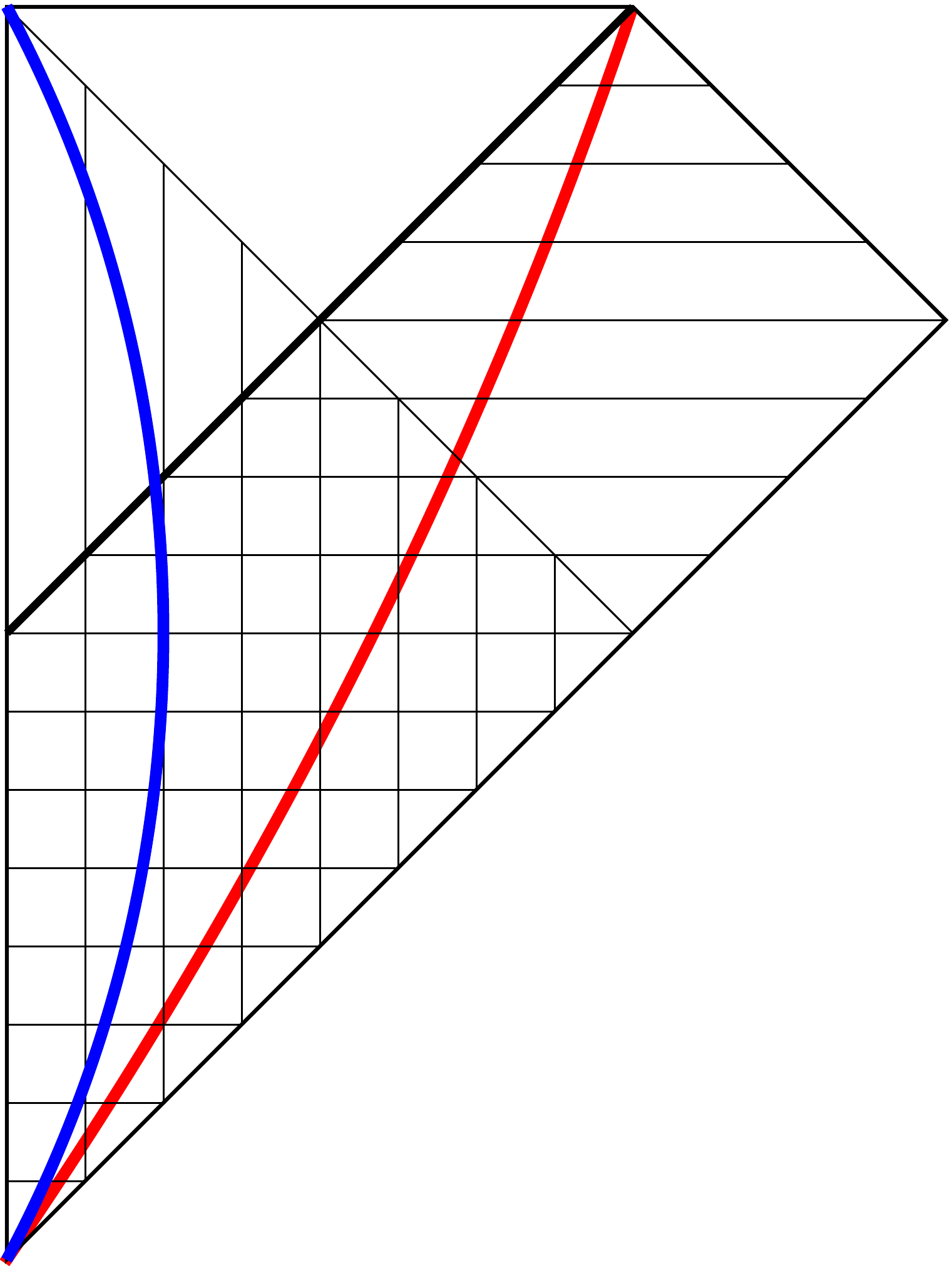}
\end{center}
\caption{A black hole formed by gravitational collapse.  The horizonal shaded region is the causal diamond of an observer (red) who remains forever outside the black hole.  The vertical shaded region is the causal diamond of an observer (blue) who falls into the black hole.  Note that there are infinitely many inequivalent infalling observers, whose diamonds have different endpoints on the conformal boundary (the singularity inside the black hole); on the other hand, all outside observers have the same causal diamond.}
\label{fig-bh}
\end{figure*}

Can an observer outside the horizon of a black hole recover information about the interior?   Consider a black hole that forms by the gravitational collapse of a star in asymptotically flat space.  Fig.~\ref{fig-bh})
shows the spacetime region that can be probed by an infalling observer, and the region accessible to an observer who remains outside the horizon.  At late times the two observers are out of causal contact, but in the remote past their causal diamonds have considerable overlap.

\begin{figure*}
\begin{center}
\includegraphics[scale = .45]{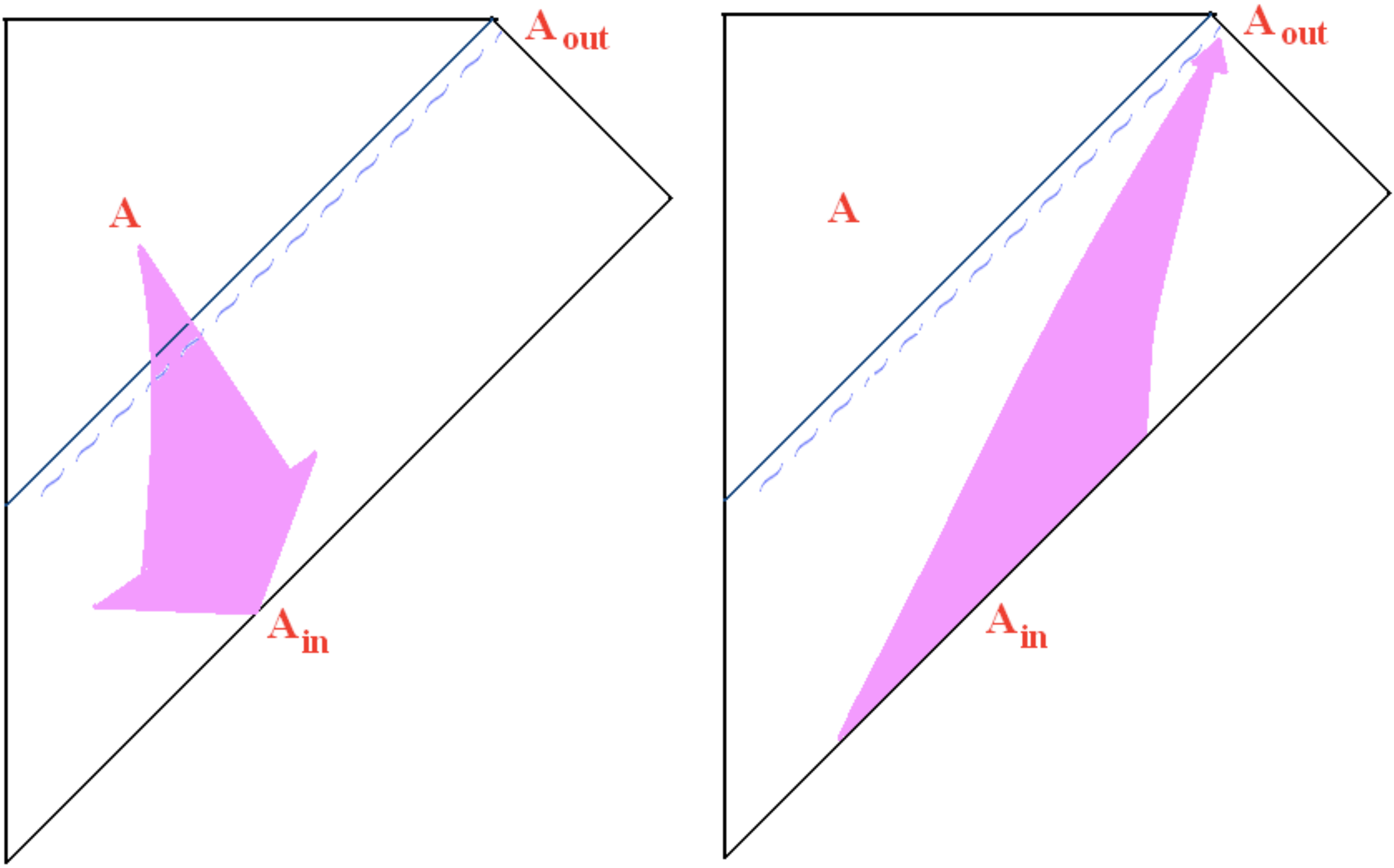}
\end{center}
\caption{An observable $A$ inside the black hole can be approximately defined and evolved using local field theory.  In this way, it can be propagated back out of the black hole to an operator $A_{\rm in}$ defined on the asymptotic past boundary.  From there, it is related by an exact S-matrix to an operator $A_{\rm out}$ that can be measured in the Hawking radiation (wiggly dashed line) by a late time observer.}
\label{fig-baf}
\end{figure*}

Let $A$ be an observable behind the horizon as shown in Fig.~\ref{fig-baf}.
$A$ might be a slightly smeared field operator or product of such field operators. To the freely falling observer the observable $A$ is a low energy operator that can be described by conventional physics, for example quantum electrodynamics. 

The question we want to ask is whether there is an operator outside the horizon on future light-like infinity, that has the same information as $A$. Call it $A_{\rm out}.$ By that we mean an operator in the Hilbert space  of the outgoing Hawking radiation that can be measured by the outside observer, and that has the same probability distribution as the original operator $A$ when measured by the in-falling observer.

First, we will show that there is an operator in the remote past, $A_{\rm in}$ that has the same probability distribution as $A$. We work in the causal diamond of the infalling observer, in which all of the evolution leading to $A$ is low energy physics. Consider an arbitrary foliation of the infalling causal diamond into Cauchy surfaces, and let each slice be labeled by a coordinate $t$.  We may choose $t=0$ on the slice containing $A$. 

Let $U(t)$ be the Schr\"odinger picture time-evolution operator, and let $|\Psi(t)\rangle$ be the state on the Cauchy surface $t$.  We can write  $|\Psi(0)\rangle$ in terms of the state at a time $-T$ in the remote past,
\begin{equation}
|\Psi(0)\rangle = U(T)|\Psi(-T)\rangle~.
\end{equation}
The expectation value of $A$ can be written in terms of this early-time state as
\begin{equation}
\langle \Psi(-T)| U^{\dag}(T)   A U(T)|\Psi(-T)\rangle~.
\end{equation}
Thus the operator
\begin{equation}
A_{\rm in}=U^{\dag}(T)   A U(T)
\end{equation}
has the same expectation value as $A$. More generally, the entire probability distributions for $A$ and $A_{\rm in}$ are the same.  Let us take the limit $T\to \infty$ so that $A_{\rm in}$ becomes an operator in the  Hilbert space of incoming scattering states. 

Since the two diamonds overlap in the remote past, $A_{\rm in}$ may also be thought of as an operator in the space of states of the outside observer.  Now let us run the operator forward in time by the same trick, except working in the causal diamond of the outside observer. The connection between incoming and outgoing scattering states is through the S-matrix. Thus we define
\begin{equation}
A_{\rm out} = S A_{\rm in} S^{\dag}
\end{equation}
or
\begin{equation}
A_{\rm out} =\lim_{T\to \infty} S U^{\dag}(T)  A U(T) S^{\dag}
\label{eq-aout}
\end{equation}
The operator $A_{\rm out}$, when measured by an observer at asymptotically late time, has the same statistical properties as $A$ if measured behind the horizon at time zero.

The low energy time development operator $U(T)$ is relatively easy to compute, since it is determined by integrating the equations of motion of a conventional low energy system such as QED. However, this part of the calculation will not be completely precise, because it involves states in the interior if the black hole, which have finite entropy bound.  The S-matrix should have a completely precise definition but is hard to compute in practice. Information that falls onto the horizon is radiated back out in a completely scrambled form. The black hole horizon is the most efficient scrambler of any system in nature.

This is the content of black hole complementarity: observables behind the horizon are not independent variables. They are related to observables in front of the horizon by unitary transformation. The transformation matrix is $\lim_{T\to \infty}[U(T) S^{\dag}]$.  It is probably not useful to say that measuring $A_{\rm out}$ tells us what happened behind the horizon~\cite{BouFre06a}. It is not operationally possible to check whether a measurement of $A$ and $A_{\rm out}$ agree. It is enough for us that the every (approximate) observable behind the horizon has a (precise) complementary image among the degrees of freedom of the Hawking radiation that preserves expectation values and probabilities.

What is the most general form of operators $A$ inside the black hole that can be written in the form of Eq.~(\ref{eq-aout}), as an operator $A_{\rm out}$ on the outside?  Naively, we might say any operator with support inside the black hole can be so represented, since any operator can be evolved back to the asymptotic past using local field theory.  But this method is not completely exact, and we know that there must be situations where it breaks down completely.  For example, by the same argument we would be free to consider operators with support both in the Hawking radiation and in the collapsing star and evolve them back; this would lead us to conclude that either information was xeroxed or lost to the outside.  This paradox is what led to the proposal that only operators with support inside someone's causal patch make any sense.  But that conclusion has to apply whether we are inside or outside the black hole; the infalling observer is not excepted.  For example, we should not be allowed to consider operators at large spacelike separation near the future singularity of the black hole.  The semiclassical evolution back to the asymptotic past must be totally unreliable in this case.\footnote{The notion of causality itself may become approximate inside the black hole.  However, this does not give us licence to consider operators at large spacelike separation inside the black hole.   Large black holes contain spatial regions that are at arbitrarily low curvature and are not contained inside any single causal patch.  By the equivalence principle, if we were permitted to violate the restriction to causal patches inside a black hole, then we would have to be allowed to violate it in any spacetime.}

We conclude that there are infinitely many inequivalent infalling observers, with different endpoints on the black hole singularity.  Approximate observables inside the black hole must have the property that they can be represented by an operator with support within the causal diamond of {\em some} infalling observer.  Any such operator can be represented as an (exact) observable of the outside observer, i.e., as an operator $A_{\rm out}$ acting on the Hawking radiation on the outside. 

This completes our discussion of black hole complementarity.   Following Refs.~\cite{FreSus04,SekSus09}, we will now conjecture a similar relation for the multiverse.  The role of the observer who remains outside the black hole will be played by the Census Taker; note that both have causal diamonds with infinite area.   The role of the many inequivalent infalling observers will be played by the causal diamonds that end in crunches, which we considered in Sec.~\ref{sec-patch}, and which have finite area.  The conjecture is

\paragraph{Hat Complementarity}  Any (necessarily approximate) observable in the finite causal diamonds of the multiverse can be represented by an exact observable in the Census Taker's hat.\\  

\begin{figure*}
\begin{center}
\includegraphics[scale = .31]{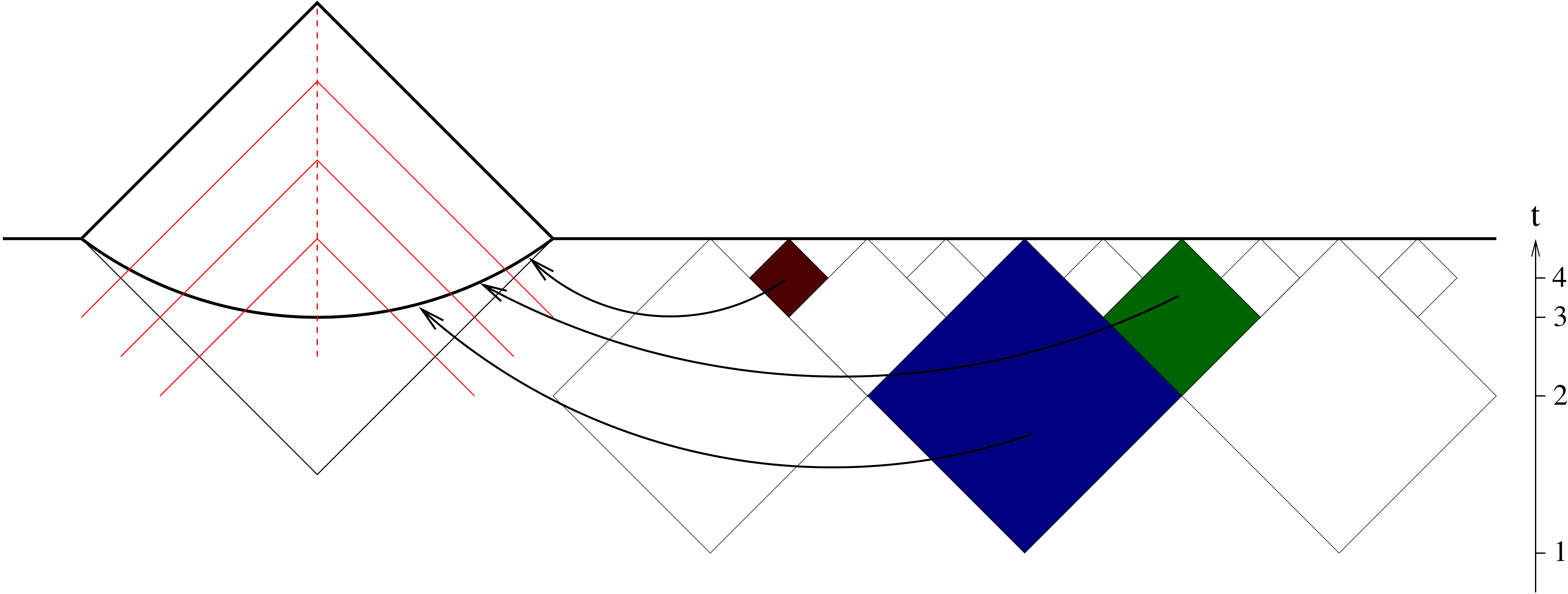}
\end{center}
\caption{Hat complementarity states that operators in finite causal diamonds (right) have an exact counterpart in the hat.  Over time, the census taker receives an unbounded amount of information.  The past light-cones of the Census Taker at different times along his worldline define a natural cutoff on his information.   Because there are only finitely many different diamonds, there must be infinitely many copies of every piece of information in the hat.  They are related by hat complementarity to the infinite number of causal diamonds that make up the global spacetime; see also Fig.~8. At late times, increasing the Census Taker's cutoff is related to increasing the light-cone time cutoff, which adds a new (redundant) layer of diamonds.}
\label{fig-hatcomp}
\end{figure*}

\noindent More precisely, we assume that for every observable $A$ in a finite causal diamond, there exists an operator $A_{\rm hat}$ with the same statistics as $A$.  $A$ and $A_{\rm hat}$ are related the same way that $A$ and $A_{\rm out}$ are in the black hole case.  See Fig.~\ref{fig-hatcomp} for a schematic illustration.

Hat complementarity is motivated by black hole complementarity, but it does not follow from it rigorously.  The settings differ in some respect.  For example, a black hole horizon is a quasi-static Killing horizon, whereas the Census Taker's causal horizon rapidly grows to infinite area and then becomes part of the conformal boundary.  Correspondingly, the radiation in the hat need not be approximately thermal, unlike Hawking radiation.   And the argument that any operator behind the horizon can be evolved back into the outside observer's past has no obvious analogue for the Census Taker, since his horizon can have arbitrarily small area and hence contains very little information at early times; and since the global multiverse will generally contain regions which are not in the causal future of the causal past of any Census Taker.  Here, we adopt hat complementarity as a conjecture.

Next, we turn to the question of {\em how} the information about finite causal diamonds shows up in the precise description of the outside observer or the Census Taker.

\subsection{The global multiverse in a hat}
\label{sec-ct}

An important difference between the black hole and the multiverse is that the information in the Hawking radiation is finite, whereas the number of particles in the hat is infinite.  The observer outside the black hole, therefore, receives just enough information to be able to recover the initial state, and thus the approximate observables inside the black hole.  On the other hand, the $O(3,1)$ symmetry of the FRW universe implies that the entropy accessible to the Census Taker is infinite.  Since the number of quantum states in the finite causal diamonds that end in crunches is bounded, this implies an infinite redundancy in the Census Taker's information.  We will now argue that this redundancy is related to the reconstruction of a global multiverse from causal diamonds, described in Sec.~\ref{sec-everett}, in which an infinite number of finite causal diamonds are used to build the global geometry.

Black holes scramble information.  Therefore, an observer outside a black hole has to wait until half the black hole has evaporated before the first bit of information can be decoded in the Hawking radiation~\cite{Pag93,SusTho93}.  After the whole black hole has evaporated, the outside observer has all the information, and his information does not increase further.  Can we make an analogous statement about the Census Taker?

If the radiation visible to the Census Taker really is complementary to the causal diamonds in the rest of the multiverse, it will surely be in a similarly scrambled form.  For hat complementarity to be operationally meaningful, this information must be organized in a ``friendly'' way, i.e., not maximally scrambled over an infinite-dimensional Hilbert space.  Otherwise, the census taker would have to wait infinitely long to extract information about any causal patch outside his horizon.\footnote{We are thus claiming that things {\em happen} in the past light-cone of the Census Taken at finite time. The corresponding irreversible entanglement leading to decoherence must be that between the interior and the exterior of the Census Taker's past light-cone on a suitable time-slice such as that shown in Fig.~\ref{fig-hatcomp}.  Because the timeslice is an infinite FRW universe, the environment is always infinite.  The size of the ``system'', on the other hand, grows without bound at late times, allowing for an infinite number of decoherent events to take place in the Census Taker's past.  Thus the postulates of Sec.~\ref{sec-sagredo} can be satisfied.}  This would be analogous to the impossibility of extracting information from less than half of the Hawking radiation.   An example of friendly packaging of information is not hard to come by (see also~\cite{HayPre07}).  Imagine forever feeding a black hole with information at the same average rate that it evaporates. Since the entropy of the black hole is bounded it can never accumulate more than $S$ bits of information. Any entering bit will be emitted in a finite time even if the total number of emitted photons is infinite.  

We will assume that this is also the case in the hat.  Then we can ask whether the Census Taker's cutoff translates to a multiverse cutoff, hopefully something like the light-cone time cutoff of Sec.~\ref{sec-dual}.

To understand the Census Taker's cutoff, we start with the metric of open FRW space with a vanishing cosmological constant. We assume the FRW bubble nucleated from some ancestor de Sitter vacuum.
\begin{equation}
ds^2 = a(T)^2 ( -dT^2 + d{\cal H}_3^2)
\end{equation}
where ${\cal H}_3$ is the unit hyperbolic geometry (Euclidean AdS)
\begin{equation}
d{\cal H}_3^2 = dR^2 + \sinh^2 {R} \ d\Omega_2^2,
\end{equation}
and $T$ is conformal time.  The spatial hypersurfaces are homogeneous with symmetry $O(3,1)$ which acts on the two-dimensional boundary as special conformal transformations. Matter in the hat fills the noncompact spatial slices uniformly and therefore carries an infinite entropy.

If there is no period of slow-roll inflation in the hat then the density of photons will be about one per comoving volume. We take the Census Taker to be a comoving worldline (see Fig.~\ref{fig-hatcomp}).  The number of photons in the Census Taker's causal past is 
\begin{equation}
N_{\gamma} \sim e^{2T_{CT}}.
\end{equation}
In this formula $T_{CT}$ is the conformal time from which the Census Taker looks back. $N_{\gamma}$ represents the maximum number of photons that the Census Taker can detect by the time $T_{CT}.$

The Census Taker's cutoff is an information cutoff: after time $T_{CT}$ a diligent Census Taker can have gathered about $e^{2T_{CT}}$ bits of information. If the de Sitter entropy of the ancestor is $S_a$, then after a conformal time $T_{CT} \sim \log S_a$ the Census Taker will have accessed an amount of information equal to the entropy of the causal patch of the ancestor. Any information gathered after that must be about causal diamonds in the rest of the multiverse, in the sense that it concerns operators like $A$ that are beyond the horizon.  

Over time, the Census Taker  receives an unbounded amount of information, larger than the entropy bound on any of the finite causal diamonds beyond the hat.  This means that the Census Taker will receive information about each patch history over and over again, redundantly.  This is reminiscent of the fact that in our reconstruction of the  global picture, every type of causal diamond history occurs over and over again as the new diamonds are inserted in between the old ones.  This is obvious because we used infinitely many diamonds to cover the global geometry, and there are only finitely many histories that end in a crunch or on an eternal endpoint.  

A particular history of a causal diamond has a larger or smaller maximum area, not a fixed amount of information.  But in our reconstruction of the global multiverse, each generation of new causal diamonds (the set of diamonds starting at the green squares at equal time $t$ in Fig.~\ref{fig-patches}) contains all possible histories (at least for sufficiently late generations, where the number of diamonds is large).  Therefore the amount of information in the reconstructed global geometry grows very simply, like $2^t$.  This redundant information should show up in the hat organized in the same manner.  Thus, it is natural to conjecture that the information cutoff of the Census Taker is dual, by complementarity, to the discretized light-cone time cutoff implicit in our reconstruction of a global spacetime from finite causal diamonds.

\acknowledgments We would like to thank T.~Banks, B.~Freivogel, A.~Guth, D.~Harlow, P.~Hayden, S.~Leichenauer, V.~Rosenhaus, S.~Shenker, D.~Stanford, E.~Witten, and I.~Yang for helpful discussions.  This work was supported by the Berkeley Center for Theoretical Physics, by the National Science Foundation (award numbers 0855653 and 0756174), by fqxi grant RFP2-08-06, and by the US Department of Energy under Contract DE-AC02-05CH11231.

\bibliographystyle{utcaps}
\bibliography{all}
\end{document}